%% file: main.tex
\documentclass[a4paper,11pt]{article}
\usepackage{pos}

\usepackage{amsfonts} % AMS
\usepackage{amssymb} % AMS
\usepackage{amsmath} % AMS
\usepackage{graphicx} % Include figure files
\usepackage{subfigure} % Include figure files
\usepackage{array} % array
\usepackage{dcolumn} % Align table columns on decimal point
\usepackage{bm} % bold math
\usepackage{latexsym} % latex symbols
\usepackage{longtable} % long tables
\usepackage{hyperref} % hypertext links
\usepackage{bm}
\usepackage{slashed}
\usepackage{bbold}
\usepackage{color}
\usepackage{multirow}
\usepackage{rotating}
\usepackage{comment}
\usepackage [english]{babel}
\usepackage [autostyle, english = american]{csquotes}
\MakeOuterQuote{"}
\usepackage{graphicx}
\usepackage{subfigure}
\usepackage{xcolor,cancel}
\usepackage{appendix}[toc,page]
\usepackage{setspace}
\usepackage{amsmath, amssymb}
\usepackage{slashed}

\newcommand{\be}{\begin{eqnarray}}
\newcommand{\ee}{\end{eqnarray}}

\providecommand{\abs}[1]{\lvert#1\rvert}
\providecommand{\matrixe}[3]{\langle#1\lvert#2\rvert#3\rangle}

\include{macro}

\title{Isovector Axial Charge and Form Factors of Nucleons from Lattice QCD}
\ShortTitle{Axial vector form factors}

\author*[a]{Rajan Gupta}

\affiliation[a]{Los Alamos National Laboratory, Theoretical Division T-2, Los Alamos, NM 87545, USA}
\emailAdd{rajan@lanl.gov}

\abstract{ I present an overview of the calculations of the isovector
  axial vector form factor of the nucleon, $G_A(Q^2)$, using lattice QCD.
  Based on a comparison of results from various collaborations, a case
  is made that lattice results are now consistent within 10\%. A
  similar level of uncertainty is found also in the axial charge
  $g_A^{u-d}$, the mean squared axial charge radius, $\langle r_A^2
  \rangle$, the induced pseudoscalar charge $g_P^\ast$, and the
  pion-nucleon coupling $g_{\pi NN}$. These lattice results for
  $G_A(Q^2)$ are already compatible with those obtained from the
  recent MINER$\nu$A experiment but lie 2-3$\sigma$ higher than the
  phenomenological extraction from the old $\nu$-deuterium bubble
  chamber scattering data for $Q^2 > 0.3$~GeV${}^2$. Fits to our data
  show that the dipole ansatz does not have enough parameters to
  parameterize the form factor over the range $0 \le Q^2 \le
  1$~GeV${}^2$, whereas even a $z^2$ truncation of the $z$-expansion
  or a low order Pad\'e are sufficient. Looking ahead, lattice QCD
  calculations will provide increasingly precise results over the
  range $0 \le Q^2 \lesssim 1$~GeV${}^2$, and MINER$\nu$A-like
  experiments will extend the range to $Q^2 \sim 2$~GeV${}^2$ or
  higher.  To increase precision of lattice data to the percent level,
  new developments are needed to address two related issues: the
  exponentially falling signal-to-noise ratio in all nucleon
  correlation functions and removing excited state contributions.
  Nevertheless, even with the current methodology, significant
  reduction in errors is expected over the next few years with higher
  statistics data on more ensembles closer to the physical point.
  \looseness-1}

\FullConference{%
 The 40th International Symposium on Lattice Field Theory (LATTICE2023)\\
  July 31st - August 4th, 2023\\
  Fermi National Accelerator Laboratory
}

\begin{document}
\maketitle

\section{Introduction}   %S01
\label{sec:intro}

The axial charge, $g_A^{u-d}$, gives the strength of the coupling of
the weak current to the nucleons. It has been determined very
accurately from the asymmetry parameter $A$ (relative to the
plane defined by the directions
of the neutron spin and the emitted electron) in the decay distribution of
the neutron, $n \to p + e^- + {\overline \nu}_e$.  The best
determination of the ratio of the axial to the vector charge,
$g_A/g_V$, comes from using (i) polarized ultracold neutrons (UCN) by
the UCNA collaboration,
$1.2772(20)$~\cite{Mendenhall:2012tz,Brown:2017mhw}, and (ii) cold
neutron beam by PERKEO III,
$1.27641(45)(33)$~\cite{Markisch:2018ndu,Mund:2012fq}. Note that, in
the SM, $g_V=1$ up to second order corrections in isospin
breaking~\cite{Ademollo:1964sr,Donoghue:1990ti} as a result of the
conservation of the vector current.

The axial charge enters in many analyses of nucleon structure, of the Standard
Model (SM), and in probes of beyond-the-SM (BSM)
physics~\cite{Bhattacharya:2011qm,Ivanov:2014bya}. For example, it
enters in the relation between the Cabibbo-Kobayashi-Maskawa (CKM)
matrix element $V_{ud}$ and the neutron lifetime, $\tau_n$. High precision extraction of 
$V_{ud}$, knowing $\tau_n$ and $g_A$, is important for 
the test of the unitarity of the first row of the CKM
matrix~\cite{Czarnecki:2018okw,Czarnecki:2019mwq,Czarnecki:2019iwz}. It
is needed in the analysis of neutrinoless double-beta
decay~\cite{Horoi:2018fls} and in the rate of proton-proton
fusion~\cite{Carroll2007}, the first step in the thermonuclear
reaction chains that power low-mass hydrogen-burning stars like the
sun.

The axial-vector form factor (AVFF) gives the dependence of this
coupling on the momentum squared transferred by the weak current to
the nucleon. It is an input in the theoretical calculation of the
neutrino-nuclei scattering cross-section needed for the analysis of
neutrino oscillation
experiments~\cite{Ruso:2022qes,Kronfeld:2019nfb,Meyer:2022mix}. The
cleanest experimental measurement would be from scattering neutrinos
off liquid hydrogen targets, however, these are not being carried out
due to safety concerns. Extractions from ongoing neutrino scattering
experiments (T2K, NOvA, MINERvA, MicroBooNE, SBN) have uncertainty due
to those in the cross-section and the incoming flux, and from the lack
of precise reconstruction of the final state of the struck
nucleus. Conversely, uncertainty in the AVFF feeds into the uncertainty in
the incoming neutrino energy that is needed to determine the
oscillation parameters.

The MINER$\nu$A experiment~\cite{MINERvA:2023avz}, using a separation
based on kinematics, has recently extracted the axial-vector form
factor of the nucleon from the charged current elastic scattering
process ${\overline{\nu}}_\mu H \to \mu^+ n$ in which the free proton
in the hydrogen (H) (part of the hydrocarbon in the target) gets
converted into a neutron. This opens the door to direct measurements
of the nucleon axial-vector form factor without the need for
extraction from scattering off nuclei, whose analysis involves nuclear
corrections that have unresolved systematics. On the theoretical
front, lattice QCD provides the best method for first principal
non-perturbative predictions with control over all sources of
uncertainty~\cite{Kronfeld:2019nfb,Ruso:2022qes}.\looseness-1

A recent comparison~\cite{Tomalak:2023pdi} of results for the AVFF
from lattice QCD~\cite{Jang:2023zts}, the MINER$\nu$A
experiment~\cite{MINERvA:2023avz}, and the phenomenological extraction
from neutrino-deuterium data~\cite{Meyer:2016oeg} showed that in the
near term the best prospects for determining the AVFF will be a
combination of lattice QCD calculations and MINER$\nu$A-like
experiments.  Lattice QCD will provide the best estimates for $Q^2
\lesssim 0.5$~GeV${}^2$, and be competitive with MINER$\nu$A for $0.5
\lesssim Q^2 \lesssim 1$~GeV${}^2$. For $Q^2 \gtrsim 1$~GeV${}^2$, new
ideas are needed for robust predictions using lattice QCD.

The goal of theory efforts in support of neutrino oscillation
experiments is robust calculations of the cross-section for targets,
such as ${}^{12}C$, ${}^{16}O$, and ${}^{40}Ar$, being used in
experiments. This involves a four step process: a precise
determination of the AVFF, nuclear models of the ground state of the
nuclei from which the neutrino scatters, the intra-nucleus evolution
of the struck nucleon using many-body theory to include complex
nuclear effects up to $\approx 5$~GeV for the DUNE experiment, and the
evolution of the final state particles to the detectors. The overall
program requires complete implementation of these within Monte Carlo
neutrino event
Generators~\cite{Ruso:2022qes,Kronfeld:2019nfb,Meyer:2022mix} with
full uncertainty quantification in each step. The output of the
generators is the essential input required by experimentalists for
determining neutrino oscillation parameters from current and future
experiments.\looseness-1

Here I review the status of lattice QCD calculations
of the axial charge, $g_A^{u-d}$ and the AVFF. In addition, note that the flavor
diagonal axial charges $g_{A}^{u,d,s,c,b}$ provide the contribution of
each quark flavor to the spin of the nucleon, whose calculation 
is computationally more expensive due to the
additional disconnected contributions.  Current status of results for these
nucleon charges has been reviewed by the Flavor Lattice Averaging
Group (FLAG) in 2019 and
2021~\cite{FlavourLatticeAveragingGroupFLAG:2021npn,FlavourLatticeAveragingGroup:2019iem}).
Including results post FLAG
2021~\cite{Alexandrou:2023qbg,Bali:2023sdi,Tsuji:2022ric,Djukanovic:2022wru,Park:2021ypf},
the values from the various calculations with 2+1- and 2+1+1-flavors
of sea quarks lie in the ranges 
$1.22 \lesssim g_A^{u-d} \lesssim 1.32$, 
$0.74 \lesssim g_A^{u} \lesssim 0.89$, 
$-0.48 \lesssim g_A^{d} \lesssim -0.38$, and 
$-0.06 \lesssim g_A^{s} \lesssim -0.025$.  There have
been no substantial new results for flavor diagonal charges since the
FLAG reports, so I will not discuss them further in this work.

Based on the results in
Refs.~\cite{Jang:2023zts,Alexandrou:2023qbg,Park:2021ypf,Djukanovic:2022wru,Alexandrou:2020okk,RQCD:2019jai},
I present the case that lattice results for AVFF over the
range $0 < Q^2 \le 1$~GeV${}^2$ are also available with $\lesssim
10\%$ uncertainty and agree with MINER$\nu$A results to within a
combined sigma as discussed in Ref.~\cite{Tomalak:2023pdi} but disagree
with the neutrino-deuterium results for $Q^2 > 0.3$~GeV${}^2$.  At the
same time, I also highlight the need for much higher statistics
and better control over excited state contributions to nucleon
correlators in lattice calculations for the uncertainty to be reduced
to the percent level.

The outline of this review is as follows.  I will summarize the
methodology and steps in the calculation of the axial and pseudoscalar
form factors in Sec.~\ref{sec:Methodology}. This includes a discussion
of the nucleon 3-point correlation functions calculated in
Sec.~\ref{sec:corr}, removing possible excited state contributions (ESC)
in Sec.~\ref{sec:ESC}, and how form factors are then obtained from them in
Sec.~\ref{sec:extractingFF}.  I then review
the operator constraint imposed on the three form factors, the axial,
$G_A(Q^2)$, the induced pseudoscalar, $\widetilde{G}_P(Q^2)$, and the
pesudoscalar $G_P(Q^2)$ by the axial Ward-Takahashi (also
referred to in literature as the partially conserved axial current
(PCAC)) identity in Sec.~\ref{sec:PCAC}, and how it provides a data driven
method for validating the enhanced contributions of multihadron, $N
\pi$, excited states. These enhanced excited state contributions are
due to the coupling of the axial and pseudoscalar currents to a pion,
i.e., the pion pole dominance hypothesis.  Extrapolation of the
lattice results to the physical point defined by the continuum ($a=0$)
and infinite volume ($M_\pi L \to \infty$) limits at physical light
quark masses in the isospin symmetric limit, i.e., $m_u = m_d$ set
using the neutral pion mass ($M_{\pi^0} = 135$~MeV) is discussed in
Sec.~\ref{sec:CCFV}. A consistency check on the extraction of the
axial charge is discussed in Sec.~\ref{sec:Acharge}.  I will then
review the results for the AVFF obtained by the various lattice
collaborations after extrapolation to the physical point in
Sec.~\ref{sec:charges}, and the comparison of lattice QCD result, the
recent MINER$\nu$A data, and the phenomenological extraction from the
old neutrino-deuterium scattering data along with my perspective on
future improvements in Sec.~\ref{sec:Minerva}. The concluding remarks are 
given in Sec.~\ref{sec:conclusions}.

\section{Calculation of the axial vector form factors using lattice QCD}   %S02
\label{sec:Methodology}

\begin{figure}[t]  %F01
\vspace{-0.5cm}
\begin{center}
\includegraphics[width=0.29\linewidth]{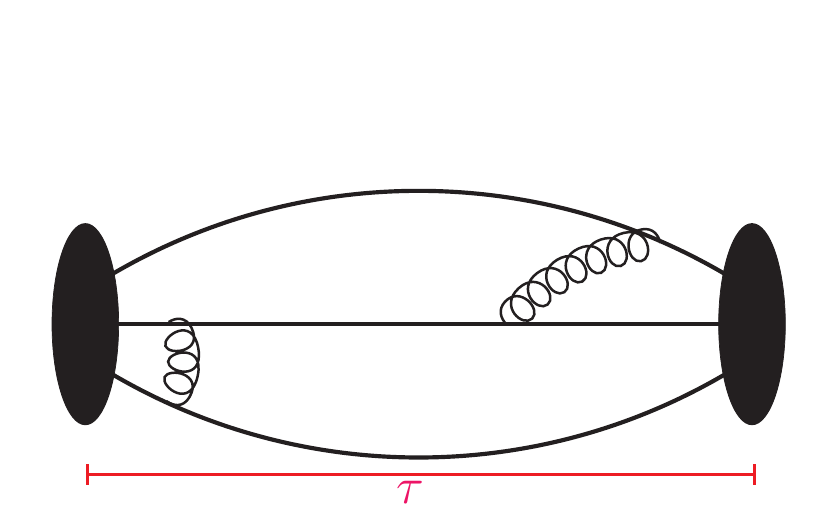}
\includegraphics[width=0.29\linewidth]{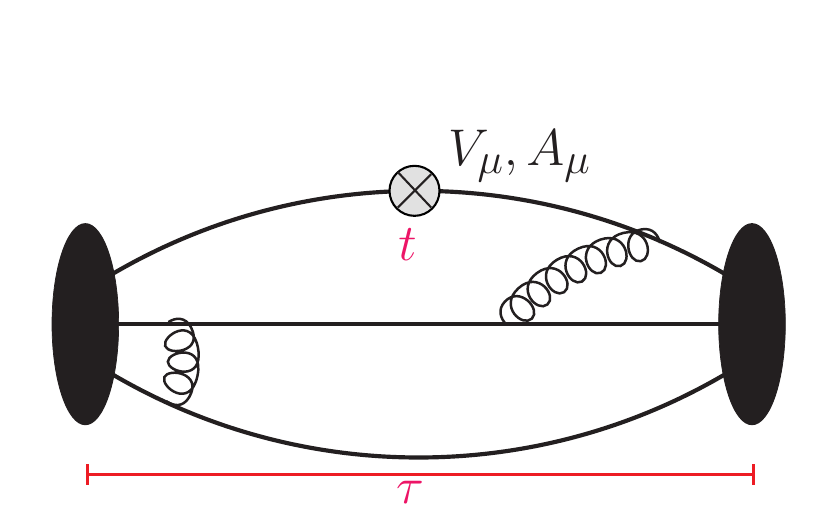}
\includegraphics[trim=70 280 0 250,clip,width=0.4\linewidth]{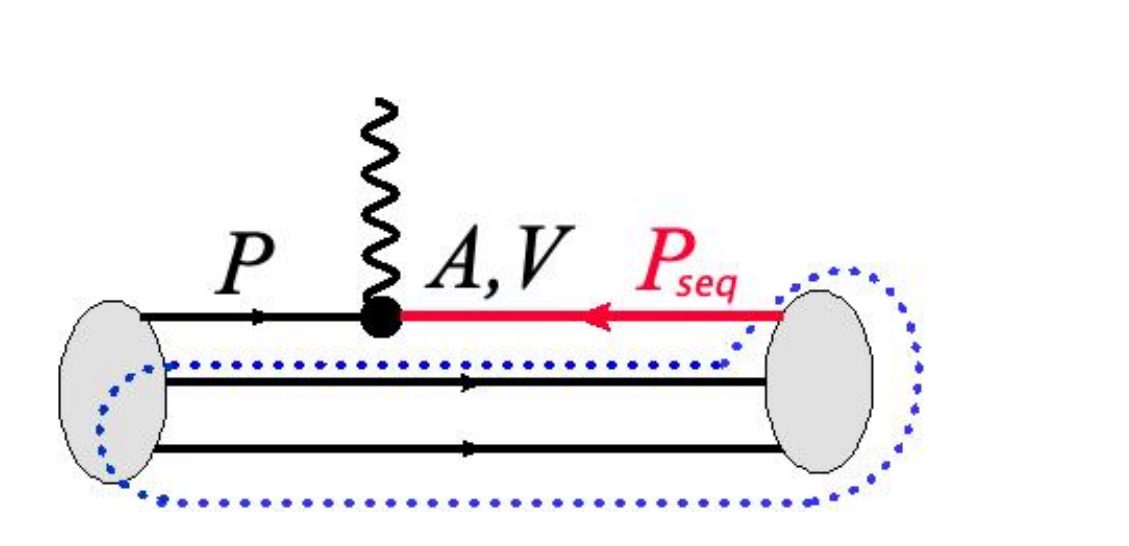}
\end{center}
\vspace{-0.5cm}
\caption{Quark line diagrams for the gauge invariant time-ordered
  correlation functions $C^\text{2pt}(\bm{p};\tau)$ and $C_J^{\rm
    3pt}(\bm{q};t,\tau)$. The gluon lines are shown only to indicate
  that all possible gluon exchanges between quarks are included, i.e.,
  it is a fully non-perturbative calculation. The electromagnetic and
  axial form factors are calculated by inserting the vector, $V_\mu$,
  and axial, $A_\mu$, currents, respectively, with momentum $\textbf
  q$ at times $t$ in between the nucleon source and sink separated by
  time $\tau$. The construction of the source for the sequential
  propagator $P_{\rm seq}$ is shown schematically in the right
  panel. The two lower quark lines within the dotted region are
  tied at the sink by $\cal N$, leaving the spin and color indices of the
  third spinor to serve as the source. \looseness-1}
\label{fig:diagrams}
\end{figure}

The quark line diagrams for the 2-point, $C^\text{2pt}$ and the
3-point $ C_J^{\rm 3pt}(\bm{q};t,\tau)$ (with the insertion of the
axial, $A_\mu$, (or vector $V_\mu$) and pseudoscalar, $P$, currents)
correlators are shown in Fig.~\ref{fig:diagrams}. The methodology for
calculating these is the same in all ongoing calculations. For $
C_J^{\rm 3pt}(\bm{q};t,\tau)$, two kinds of quark propagators are
calculated by solving, using Krylov solvers such as conjugate gradient
and accelerated using multigrid~\cite{Babich:2010qb}, the linear
equation $D P = \eta$ where $D$ is the Dirac matrix on the lattice and
$\eta$ is a source vector.  The first, $P$, is constructed using a
delta function or smeared $\eta$ and shown moving forward from the
source location, i.e., from the left blob in the right panel
in Fig.~\ref{fig:diagrams}.  The second is a sequential propagator,
$P_{\rm seq}$, shown moving backwards (red line from the right blob
representing the sink) from $\eta= nucleon$ source with definite
momentum ${\textbf p}_f$. This {\it nucleon} source, shown
schematically by the part of the diagram lying inside the dotted area,
is constructed by contracting together two original ($P$) propagators.
The insertion of the current with 3-momentum $\textbf q$ between the
source and sink nucleons then reduces to that between the original
propagator and the sequential propagator, again shown schematically by
the top line. By momentum conservation, the source nucleon is then
projected to momentum ${\textbf p}_i = {\textbf p}_f - {\textbf q} $.
The Euclidean 4-momentum transfer squared is given by $Q^2 = {\textbf
  q}^2 - (E_N - M_N)^2$.\looseness-1

In current calculations (the standard method) the nucleon interpolating operator, ${\cal N}$, used is
\begin{align}
  {\cal N}(x) =& \epsilon^{abc} \left[ q_1^{aT}(x) C \gamma_5 \frac{1\pm\gamma_4}{2} q_2^b(x) \right] q_1^{c}(x) \,,
\label{eq:Nchi}
\end{align}
where $C = \gamma_4 \gamma_2$ and the optional factor $1\pm\gamma_4$
projects on to positive parity nucleon states propagating
forward/backward in time for zero momentum correlators. Developing a variational 
basis of interpolating operators that includes all states making significant
contributions, including $N \pi$ states, i.e., the holy grail of taming ESC,
 is still work under progress~\cite{Barca:2022uhi,Grebe:2023tfx}.

A short description of the six steps in the calculation of the AVFF
that are common to all fermion discretization schemes and independent
of the selection of input simulation parameters is given next in 
Secs.~\ref{sec:corr}--\ref{sec:Acharge}. 

\subsection{Correlation functions $C^{{\rm 2pt}}$ and $ C_J^{\rm 3pt}({\textbf q};t,\tau)$}   %S02A
\label{sec:corr}

Two kinds
of smeared sources, $\eta$, have been used to generate the original and
sequential quark propagators in most lattice calculations: (i)
Wuppertal~\cite{Gusken:1989ad} and (ii) exponential
source~\cite{Tsuji:2022ric}. The quark propagators so obtained are stitched
together to construct the gauge invariant time-ordered correlation
functions $C^\text{2pt}(\bm{p};\tau)$ and $C_J^{\rm 3pt}(\bm{q};t,\tau)$ shown
in Fig.~\ref{fig:diagrams}, whose spectral decompositions are\looseness-1
\begin{align}
  C^\text{2pt}(\bm{p};\tau) \equiv& \matrixe{\Omega}{{\cal T}({\cal N}(\tau) \wbar{{\cal N}}(0))}{\Omega} = \sum_{i=0} \abs{A_i^\prime}^2 e^{-E_i \tau} \,, 
  \label{eq:C2pt} \\
  C_J^{\rm 3pt}(\bm{q};t,\tau) \equiv& \matrixe{\Omega}{{\cal T}({\cal N}(\tau) J_\Gamma (t)\wbar{{\cal N}}(0))}{\Omega} 
  = 
  \sum_{i,j=0} {A_i^\prime} {A_j^\ast} \matrixe{j}{J_\Gamma}{i^\prime} e^{-E_i t -M_j(\tau-t)} \,, 
  \label{eq:C3pt-decomp}
\end{align}
where $J_\Gamma = A_\mu = {\overline \psi} \gamma_\mu \gamma_5 \psi$
or $ J_\Gamma = P = {\overline \psi} \gamma_5 \psi$ is the quark
bilinear current inserted at time $t$ with momentum $\bm q$, and
$\ket{\Omega}$ is the vacuum state.  In the $ C_J^{\rm
  3pt}(\bm{q};t,\tau)$, using the quantum mechanical right to left
time ordering, the nucleon in the final state $\bra{j}$ is, by
construction, projected to zero momentum, i.e., $p_j = (M, \bm{0})$,
whereas the initial state $\ket{i^\prime}$ is projected onto definite
momentum $p_i = (E,\bm{p})$ with $\bm{p} = - \bm{q}$ by momentum
conservation.  The prime in $\ket{ i^\prime}$ indicates that this
state can have non-zero momentum. Consequently, the states on the two
sides of the inserted operator $J$ are different for all ${\textbf q}
\neq 0$. The goal is to extract the ground-state matrix elements
(GSME), $\matrixe{0}{J}{0^\prime}$, from fits to
Eq.~\eqref{eq:C3pt-decomp}.\looseness-1

A major challenge in the analysis of all nucleon correlators is the exponential decay of 
the signal-to-noise ratio, i.e., as $e^{-(M_N -   1.5M_\pi) \tau}$ with the source-sink separation
$\tau$~\cite{Parisi:1983ae,Lepage:1989hd}.  With 
($O(10^5)$ measurements), a good signal in $C^\text{2pt}(\bm{p};\tau)$
and $C_J^{\rm 3pt}(\bm{q};t,\tau)$ extends to $\approx 2$ and $\approx 1.5$~fm,
respectively.\looseness-1

At these $\tau$, the residual contribution of many theoretically
allowed radial and multihadron excited states are observed to be
significant. These states arise because the standard nucleon
interpolating operator ${\cal N}$, defined in Eq.~\eqref{eq:Nchi} and
used to construct the correlation functions given in
Eqs.~\eqref{eq:C2pt} and~\eqref{eq:C3pt-decomp}, couples to a nucleon
and all its excitations with positive parity including multihadron
states, the lowest of which are $N({\bf p}) \pi(-{\bf p})$
with $|{\bf p}_{\rm lowest}| =2\pi / La$ and $N({\bf 0}) \pi({\bf 0}) \pi({\bf 0})$.
Since it is not known, {\it a priori}, which excited states
contribute significantly to a given $C_J^{\rm 3pt}(\bm{q};t,\tau)$,
the first goal is to develop methods to identify these and remove
their contributions. Operationally this boils down to
knowing/determining the energies $E_i$ to put in fits to data using
the theoretically rigorous (for unitary actions) spectral
decomposition given in Eq.~\eqref{eq:C3pt-decomp}. Note that the $A_i$ are not
needed as they come in combinations ${A_i^\prime} {A_j^\ast}
\matrixe{j}{J_\Gamma}{i^\prime}$, which are fit parameters and never
used.\looseness-1

%%%%%%%%%%%%%%%%%
\begin{figure*}[htbp] %F02
\subfigure
{
    \includegraphics[width=0.235\textwidth]{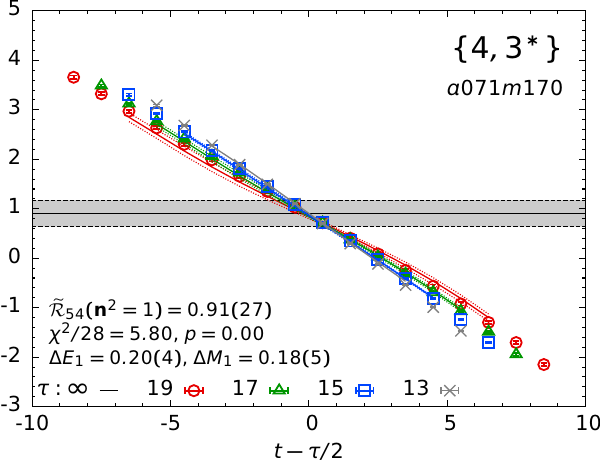}  %figs_Aesc/
    \includegraphics[width=0.235\textwidth]{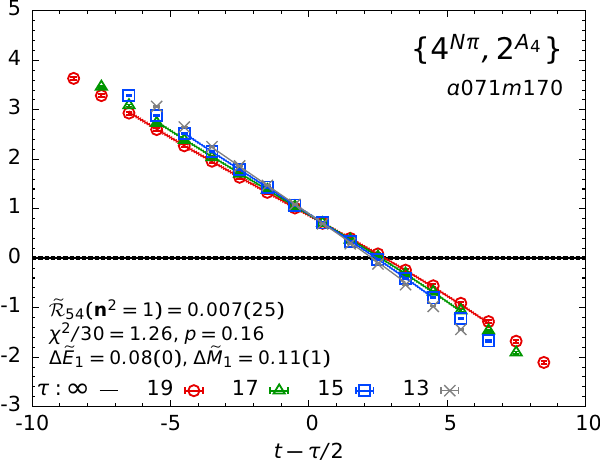} %figs_Aesc/  
}
{
    \includegraphics[width=0.23\textwidth]{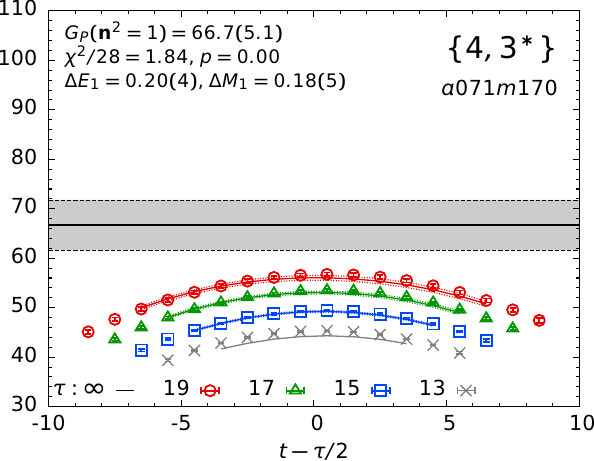} 
    \includegraphics[width=0.23\textwidth]{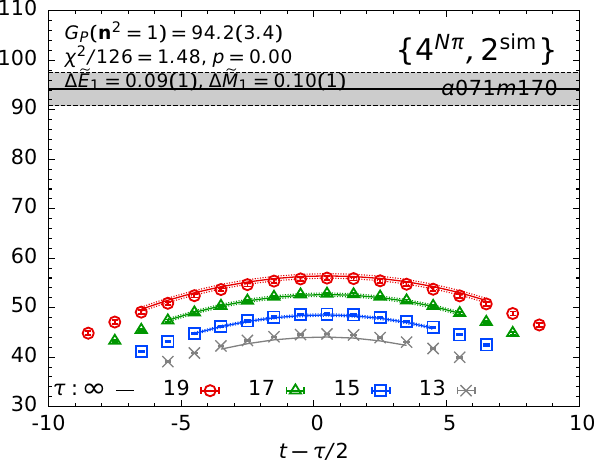} 
}
{
    \includegraphics[width=0.23\textwidth]{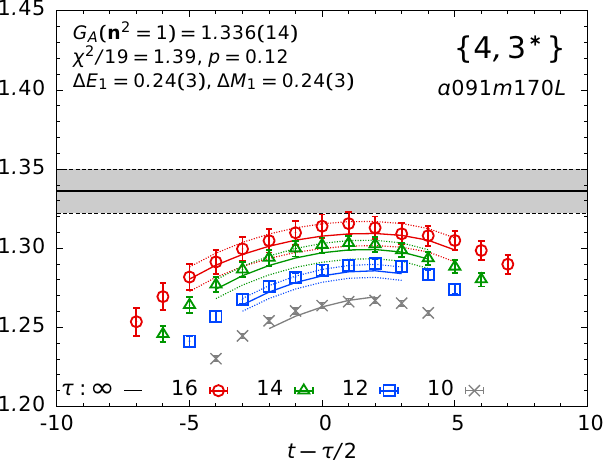}  %figs_Aesc/
    \includegraphics[width=0.23\textwidth]{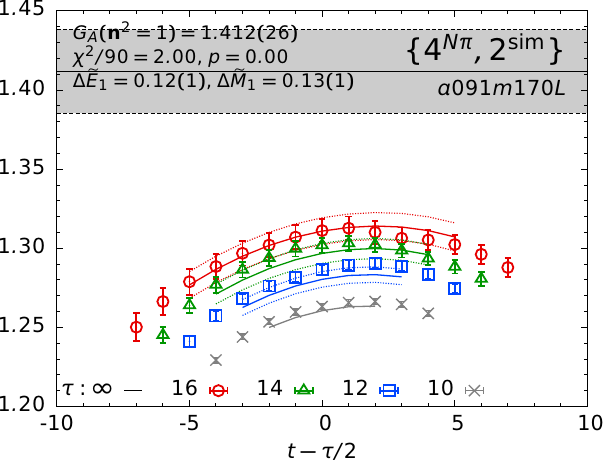}  %figs_Aesc/dup/
}
{
    \includegraphics[width=0.23\textwidth]{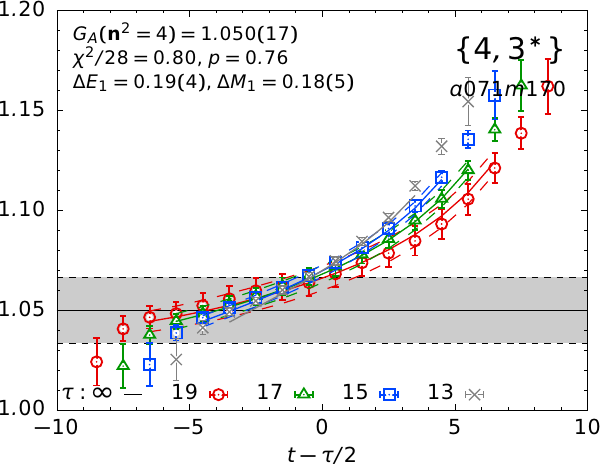}  %figs_Aesc/
    \includegraphics[width=0.23\textwidth]{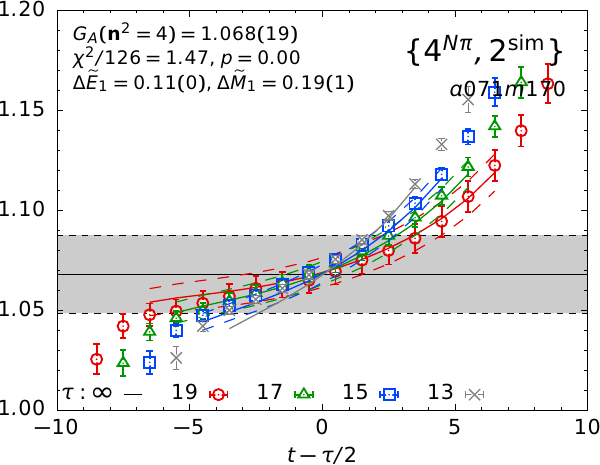}  %figs_Aesc/dup/
}
{
   \includegraphics[width=0.23\textwidth]{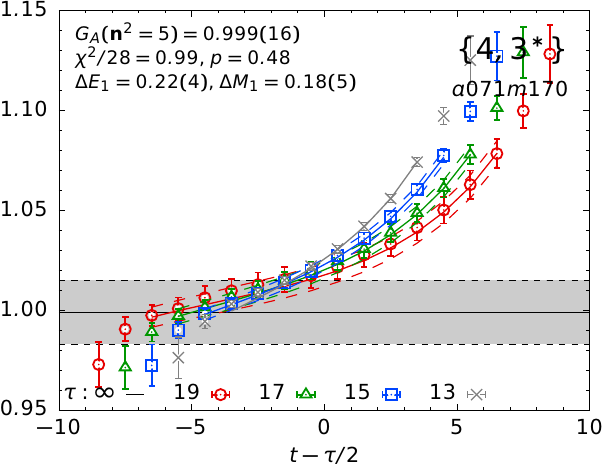}  \hspace{0.04in} %figs_Aesc/
   \includegraphics[width=0.23\textwidth]{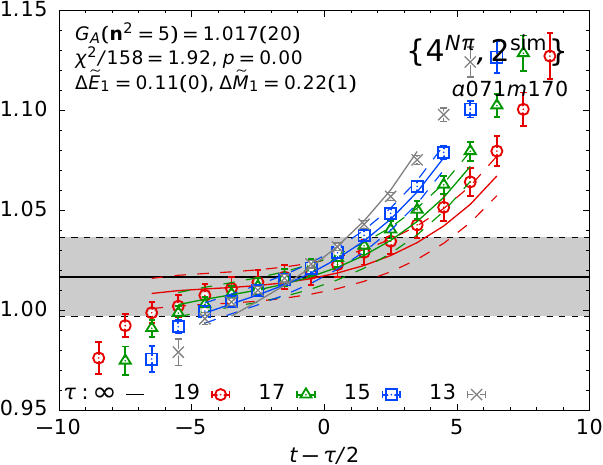} \hspace{0.09in}  %figs_Aesc/dup/
}
{
    \includegraphics[width=0.23\textwidth]{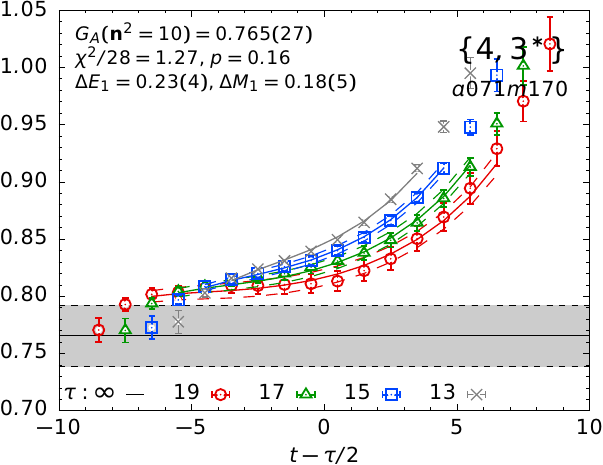} \hspace{0.04in}  %figs_Aesc/
    \includegraphics[width=0.23\textwidth]{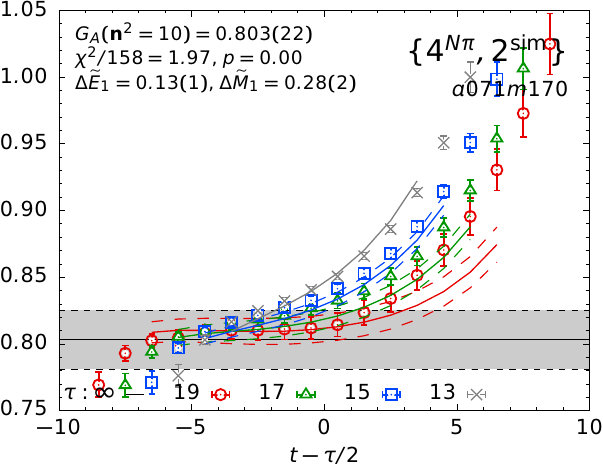}  %figs_Aesc/dup/
}
\caption{Data for the ratio $C_J^{\rm 3pt}(\bm{q},t,\tau)/\sqrt{C^{\rm 2pt}(\bm{q},t)  C^{\rm 2pt}(0,\tau -t)}$ 
 that, in the limits $(\tau -t) \to \infty$ and $t \to \infty$, should be independent of 
  $\tau$ and $t$, i.e., lie on a horizontal line in the center about  $t=\tau/2$ with value that is the GSME. Current data show large 
  ESC, and the grey band is the estimate of GSME given by the fit to Eq.~\protect\eqref{eq:C3pt-decomp}. 
  In each row, the data in each pair of panels are the same but the fit on the left is without 
  the $N \pi$ state and on the right is with. 
  The top row (panels 1 and 2) show the data and fit to $J = A_4$ with ${\bm n} = (0,0,1)$.
  These two panels illustrate (i) the improvement ($\chi^2/dof$) in the fit to $J=A_4$ data with the inclusion of 
  the $N\pi$ state and (ii) a very large ESC indicated by the large slope slowly rotating counterclockwise 
  to the expected horizontal band. 
  The right two panels show the data and fits to $J = P$ with ${\bm n} = (1,0,0)$ that should give $G_P$. 
  These two panels illustrate that the difference in $G_P$ with and without including the $N \pi$ state 
  is about $50\%$ (enhanced ESC)   and the $\chi^2/dof$ is better with the $N \pi$ state.
  Panels in rows two and three show the data and fit to $J = A_3$ with ${\bm n} = (1,0,0)$, $ (2,0,0),\ (2,1,0),\ (3,1,0)$ 
  that should give $G_A$ for $\tau \to \infty$. 
  Each pair of panels illustrates that the difference in $G_A$ with and without including the $N \pi$ state 
  is a few percent and the $\chi^2/dof$ of the two fits is comparable.
  Note the change in the behavior: the ${\bm n} = (1,0,0)$ data converge from below, while for (1,1,1) 
  and higher momenta, the data are rotating clockwise to the expected horizontal line. Also, the fits become less robust with increasing $\bm n$. 
  See Ref.~\protect\cite{Park:2021ypf} for details on these data and the fits.}
  \label{fig:ESCcomp}
\end{figure*}

\begin{figure}[ht]   %F03
\begin{center}
  \includegraphics[trim=0 0 0 0, clip, width=0.45\textwidth]{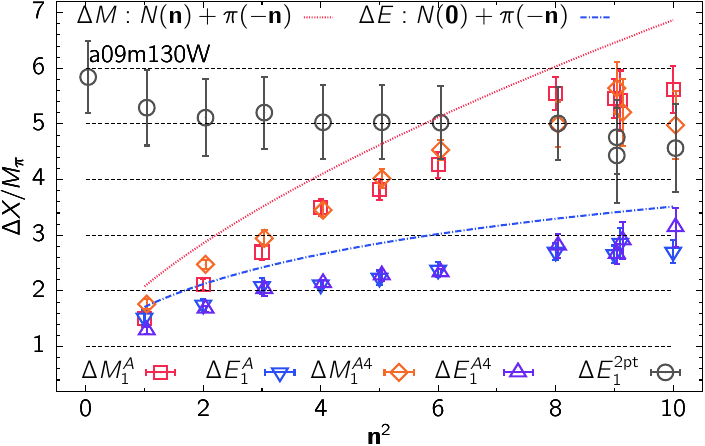} \hspace{0.2in}
  \includegraphics[trim=0 0 0 0, clip, width=0.45\textwidth]{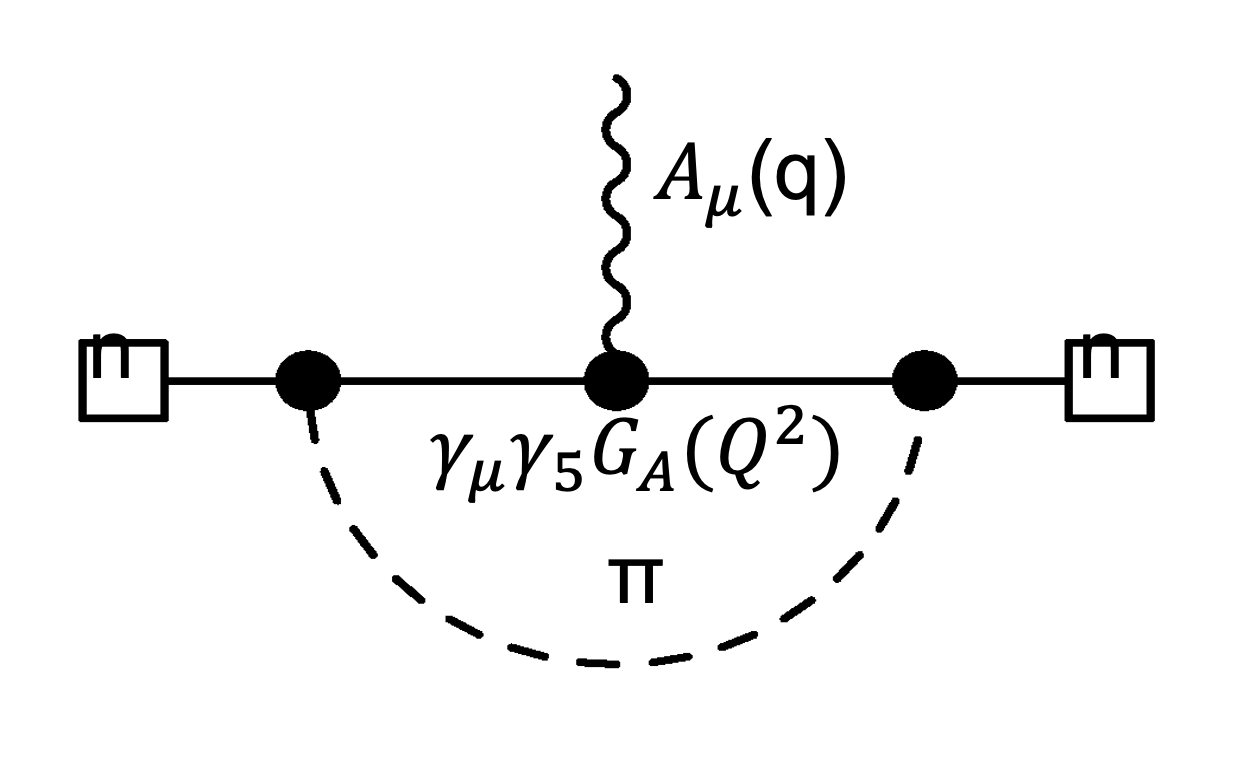}
% \label{fig:ChiPT-1-loop}
\end{center}
\vspace{-0.5cm}
  \caption{(Left) Results for the energy gaps, labeled $\Delta  E_1^{A4}$ and $\Delta M_1^{A4}$,  
    for the first excited state extracted from fits to the
    $C_{A_4}$ correlator.  These mass gaps are compared
    with the first excited state energy $\Delta E_1^{\rm 2pt}$ from
    four-state fits to the nucleon two-point correlator. Note that the
    difference between them (black circles versus blue triangles), and consequently the difference between
    the form factors extracted increases as $M_\pi \to 135$~MeV and ${\bm {n^2}} \to 0$ (equivalently $Q^2\to 0$).
    (Right) {The 1-loop correction to the 3-point function in $\chi$PT}.
}
 \label{fig:C4E1}
\end{figure}

\begin{figure}[ht]  %F04
\begin{center}
\includegraphics[width=0.35\linewidth]{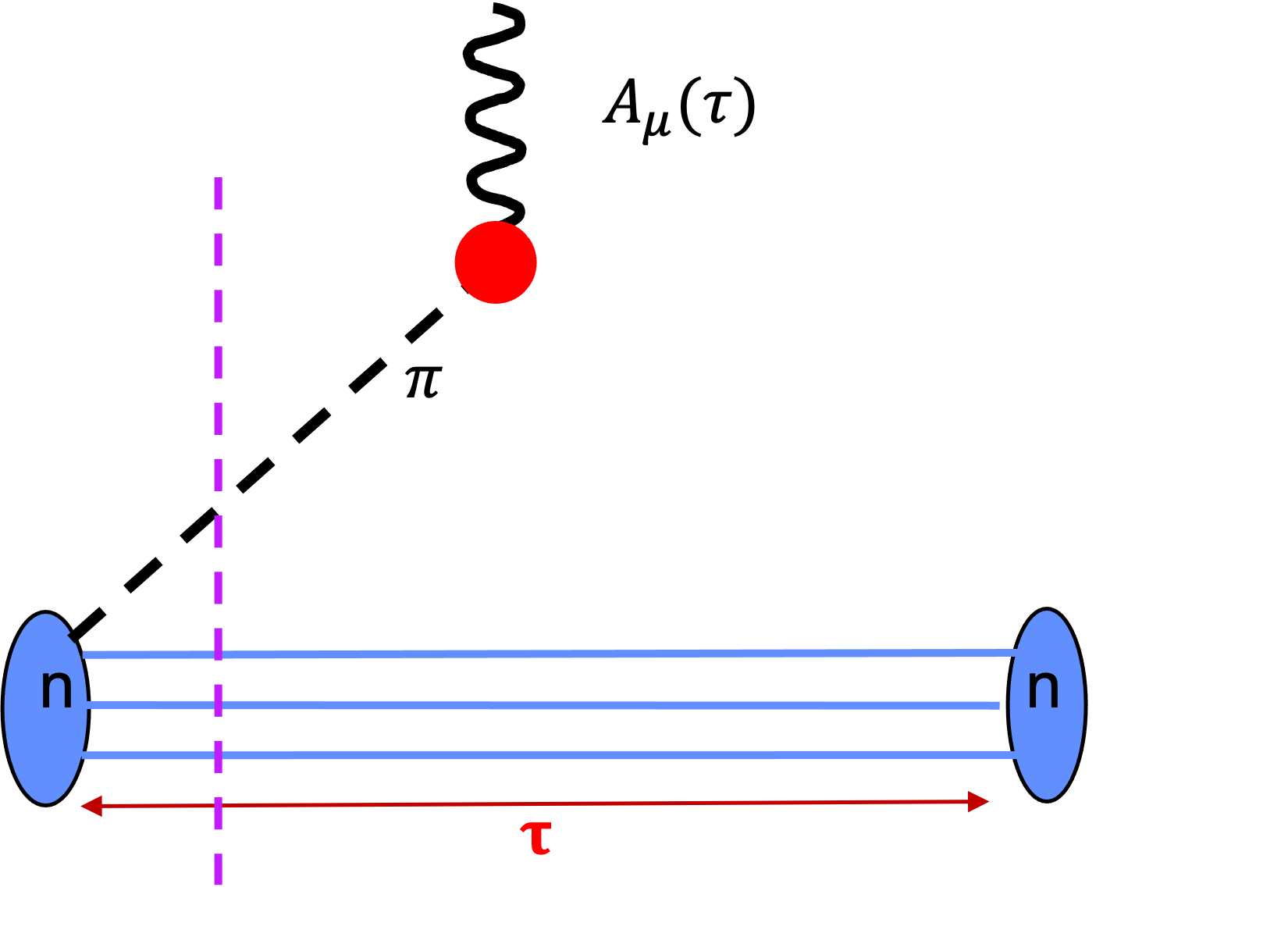}
\includegraphics[width=0.62\linewidth]{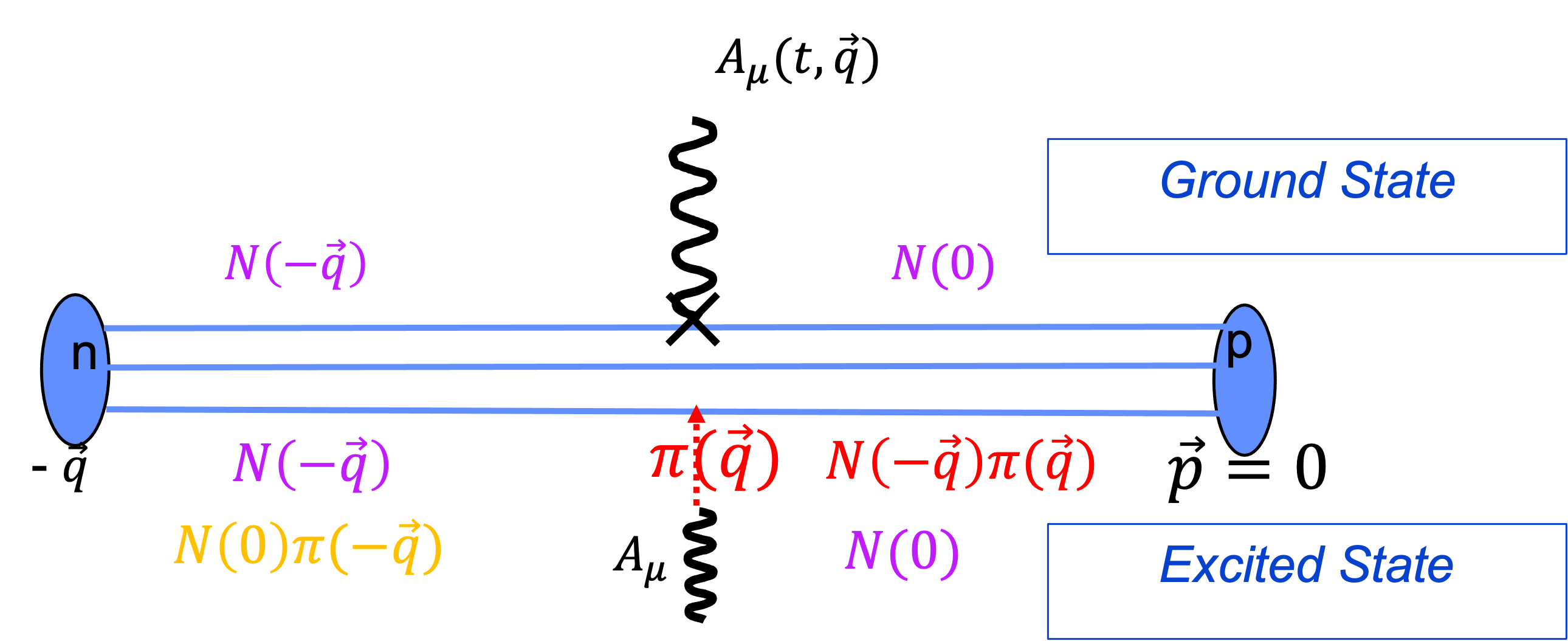}
\end{center}
\vspace{-0.5cm}
\caption{The quark line diagrams illustrating the contribution of $N \pi$ states. 
(Left) the current $A_\mu$ annihilates the pion produced by the source. (Right) 
  The states involved in the transitions: the ground state process is shown above the quark line diagram and those 
  involving an excited state on one side of the current insertion are shown below. }
\label{fig:ESC}
\end{figure}

%%%%%%%%%%%%%%%%%%%%%%%%%%%%%%%%

\subsection{Extracting the ground state matrix elements: exposing and incorporating $N \pi$ states}    %S02B
\label{sec:ESC}

The most direct way to extract $\matrixe{0}{J}{0^\prime}$ is to make
fits to Eq.~\eqref{eq:C3pt-decomp} keeping as many intermediate states
as allowed by data's precision and demonstrate convergence. The
problem is that even unconstrained 2-state fits are numerically
ill-posed.  The next option is to take the $E_i$ from
$C^\text{2pt}(\bm{p};\tau)$ as ${\cal N}$ creates the same set of
states in $C^\text{2pt}$ and $C^\text{3pt}_J$ and input these in fits
to $C_J^{\rm 3pt}(\bm{q};t,\tau)$, either by doing simultaneous fits or via
priors within say a bootstrap process to correctly propagate the
errors. Of these, the ground state ${A_0^\prime}$, ${A_0}$, and $E_0$
are well-determined from fits to the 2-point function. Similarly, one
would expect $E_1$ can also be taken from $C^\text{2pt}$.  This was
the strategy used until 2017 when it was shown in
Ref.~\cite{Rajan:2017lxk} that the resulting form factors do not
satisfy the constraint imposed on them by PCAC. Deviations from PCAC
due to discretization effects of about $\approx 5\%$ were expected,
however, almost a factor of two was found on the physical pion mass
ensembles.

The reason was provided by B\"ar~\cite{Bar:2018akl,Bar:2018xyi} using
$\chi$PT: enhanced contributions to ME from multihadron, $N \pi\,,
\ldots$, excited states that have much smaller mass gaps than of
radial excitations, the lowest being $\approx 1230$ versus
N(1440). These states were not evident in fits to
$C^\text{2pt}(\bm{p};\tau)$ as they have small amplitudes. A different
approach to analysis that includes the $N \pi$ states was needed.

It turned out that 2-state fits to $C^{\rm 3pt}_{A_4}$ exposed these
states and provided a data-driven method~\cite{Jang:2019vkm}. These
fits confirmed that the lowest of the tower of $N({\bf p}) \pi(-{\bf
  p})$ states makes a very significant contribution. By itself,
$C^{\rm 3pt}_{A_4}$ is dominated by excited states and fits to it
using the $E_i$ from $C^\text{2pt}$ gave very poor $\chi^2/dof$
. Making fits leaving $E_1$ a free parameter dramatically improved the
$\chi^2 /dof$ (compare the left two panels in the top row of
Fig.~\ref{fig:ESCcomp}), and the resulting output values of $E_1$ on
the ${\bf p}=0$ (${\bf p}$) side of operator insertion were roughly consistent
with $N({\bf p}=1) \pi({\bf p}=-1)$ ($N(0) \pi({\bf p})$) as shown in Fig.~\ref{fig:C4E1}
(left), reproduced from Ref.~~\cite{Jang:2019vkm}.  An illustration of
the current understanding of the process giving GSME
and of those involving the lowest excited states contributing is shown in
Fig.~\ref{fig:ESC} (right). The caption of Fig.~\ref{fig:ESCcomp}
points out some of the features of ESC observed in current data and
the efficacy of fits to
% the spectral decomposition of 
$C_J^{\rm 3pt}$ with and without including the lowest $N \pi$ state to
remove the ESC. \looseness-1

The fact that there is enhancement of ME in the axial channel has been
understood for over 60 years as the ``pion pole dominance'' (PPD)
hypothesis. On the lattice, the creation of a $N \pi $ state by $\cal
N$ is suppressed by $V$, the 3-d volume, compared to just the nucleon
as each state has a normalization factor of $1/V$ for a point (local)
source $\cal N$.  The axial current can, however, couple to this pion,
and because the pion is light, this coupling can occur anywhere in the
time slice at which the current is inserted with momentum $\bm q$ (see
Fig.~\ref{fig:ESC} (left)). This gives a factor of $V$ enhancement,
approximately cancelling the normalization factor
$1/V$~\cite{Bar:2018akl,Bar:2018xyi}. This enhanced contribution to
the ME when the pion comes on-shell is an artifact that has to be
removed.  Note that since energy is not conserved on the lattice, both
the neutron and the pion can come on shell, however, since momentum is
conserved, possible excited states must have the same total momentum
as the created neutron state. PPD tells us that the axial current with
momentum $\bm q$ can be viewed as the insertion of a pion with $\bm
q$, and this has a large coupling to the nucleon. These processes are
illustrated in Fig.~\ref{fig:ESC}.\looseness-1

Having identified large contributions from the $N({\bf p}=1) \pi({\bf
  p}=-1)$ state, certainly in the extraction of $\GP$ and $G_P$, the
question is---do we need to include other multihadron and radial
excited state contributions if we want results with percent level
precision?  What about in $G_A$? Note that, in addition to the
enhanced contribution shown in Fig.~\ref{fig:ESC}, $\chi$PT also
indicates that the 1-loop contributions due to the diagram shown in
Fig.~\ref{fig:C4E1} (right) (again a $N \pi$ contribution) could be
$O(5\%)$ in all the five $C_J^{\rm 3pt}$. Thus the $N\pi$ state could also be significant
for extracting $G_A$ and the charge $g_A$ (the GSME $\matrixe{0}{A_3}{0^\prime}$) from
$C_{A_3}$ at the percent precision desired. Based on these arguments, it
is clear that one needs at least 3-state fits to the five $C_J^{\rm 3pt}$ in
Eq.~\eqref{eq:C3pt-decomp}---the ground state, the $N\pi$ state and
the third that effectively accounts for all other excited state
contributions.

In my evaluation, details of the fits made to remove ESC are the most
significant differences between the calculations performed by the
different collaborations. With the current methodology, higher
statistics data are badly needed to improve these fits, i.e., include
more states in the fits, and gradually reduce the dependence on
exactly how the analyses are done.

A very important point to remember in such analyses is that fit
parameters in a truncated ansatz (the $A_i$ and $E_i$ in say a 3-state
fit in our case) try to incorporate the effects of all
contributions. Thus the connection between parameters coming out of
fits to a truncated Eq.~\eqref{eq:C3pt-decomp} and physical states
made in Figs.~\ref{fig:C4E1} and~\ref{fig:ESC} is very approximate at
best.  Thus, when I write ``$N \pi$ state'' I really only mean an
$E_1$ close to that of a non-interacting $N(1) \pi(-1)$ state and/or an $N(0)
\pi(0) \pi(0)$ state that is essentially degenerate on our lattices.

\subsection{Extracting the form factors}    %S02C
\label{sec:extractingFF}

Once the GSME, $\matrixe{0}{J}{0^\prime}$, have been extracted, their
Lorentz covariant decomposition into the axial $G_A$, induced
pseudoscalar $\widetilde{G}_P$, and pseudoscalar $G_P$ form factors is
\begin{align}
  \matrixe{N(\bm{p}_f)}{A_\mu (\bm{q})}{N(\bm{p}_i)}  &=
  {\overline u}(\bm{p}_f)\left[ G_A(Q^2) \gamma_\mu \gamma_5
  + q_\mu \gamma_5 \frac{\widetilde{G}_P(Q^2)}{2 M}\right] u(\bm{p}_i) \,,
  \label{eq:aff-a} \\
 \matrixe{N(\bm{p}_f)}{P (\bm{q})}{N(\bm{p}_i)}  &=
  {\overline u}(\bm{p}_f)\left[ G_P(Q^2) \gamma_5 \right]u(\bm{p}_i) \,,
  \label{eq:aff-ps}
\end{align}
where $u(\bm{p})$ is the nucleon spinor with momentum $\bm{p}$, $q
= p_{f} - p_{i}$ is the momentum transferred by the current, $Q^2 =
-q^2 = \bm{q}^2 - (E({\bm p}_f)-E({\bm p}_i))^2$ is the space-like
four momentum squared transferred.  On the lattice, the discrete
momenta are ${\bm p} = 2 \pi {\bm n}/La = 2 \pi (n_x,n_y,n_z)/La$ with $n_i \in \pm \{0\ldots L\}$.
The spinor normalization used is
\begin{align}
  \sum_s u(\bm{p},s) \wbar{u}(\bm{p},s) &= \frac{E(\bm{p})\gamma_4 - i\gamma\cdot\bm{p} +M}{2E(\bm{p})} \,.
\end{align}
It is important to note that the excited states have to be removed
from the correlation functions,$C$, which have the spectral decomposition given in
Eq.~\eqref{eq:C3pt-decomp}, and not from the form factors, i.e., after the
decompositions. Eqs.~\eqref{eq:aff-a} and~\eqref{eq:aff-ps} are only
valid for GSME, i.e., $\ket{N(\bm{p}_i)}$ and $\bra{N(\bm{p}_f)}$ are
assumed to be ground 
states of the nucleon. If there are residual ESC, then additional
``transition'' form factors have to be included in the rhs of
Eqs.~\eqref{eq:aff-a} and~\eqref{eq:aff-ps}.

% Extracting GSME is the key issue in getting
% high precision AVFF from lattice calculations.

Assuming GSME have been extracted, and choosing the nucleon spin projection
to be in the ``3'' direction, the explicit form of the decompositions in 
Eqs.~\eqref{eq:aff-a} and~\eqref{eq:aff-ps} become
\begin{align}
  C_{A_{\{1,2\}}}(\bm{q}) \to& K^{-1} \left[ -q_{\{1,2\}} q_3 \frac{\GP}{2M} \right]    \,,
  \label{eq:Ai-decomp} \\
  C_{A_{3}}(\bm{q}) \to& K^{-1} \left[ -q_3^2 \frac{\GP}{2M} + (M+E) G_A\right]    \,,
  \label{eq:A3-decomp} \\
  C_{A_4}(\bm{q}) \to& K^{-1} q_3 \left[(M-E)\frac{\GP}{2M} + G_A\right] \,,\\
  C_{P}(\bm{q}) \to& K^{-1} q_3 G_P \,,
\label{eq:P-decomp} 
\end{align}
where the kinematic factor $K \equiv \sqrt{2E(E+M)}$. In each case, data with all
equivalent momenta that have the same ${\bm q}^2$ are averaged to
improve the statistical signal. These correlation functions are
complex valued, and the signal, for the CP symmetric theory, is in
$\Im C_{A_i}$, $\Re C_{A_4}$, and $\Re C_{P}$.

It is clear that $G_P$ is determined uniquely from $C_P$
(Eq.~\eqref{eq:P-decomp}), and for certain momenta $\GP$ from
$C_{A_{\{1,2\}}}$ using Eq.~\eqref{eq:Ai-decomp}. The
$C_{A_{3}}(\bm{q})$ and $ C_{A_4}(\bm{q})$ give linear combinations of
$G_A$ and $\GP$, and Eq.~\eqref{eq:A3-decomp} gives only $G_A$ when 
$q_3=0$.\looseness-1

\subsection{Satisfying PCAC}     %S02D
\label{sec:PCAC}

\begin{figure}[ht]   %F06
\begin{center}
\includegraphics[width=0.49\linewidth]{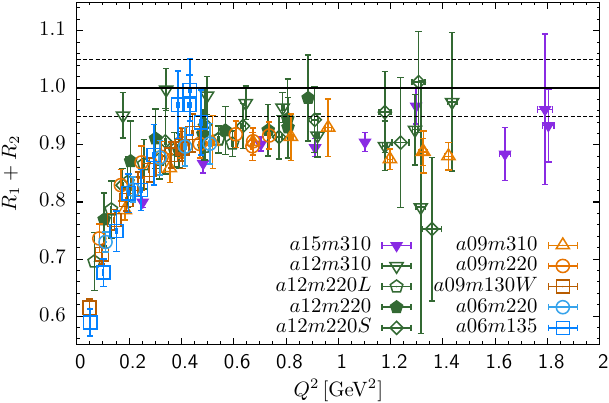}
\includegraphics[width=0.49\linewidth]{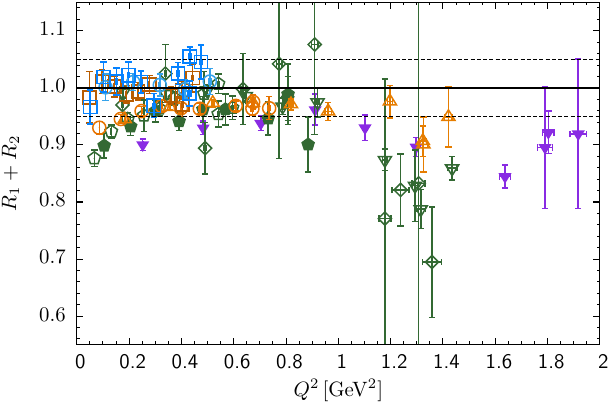}
\end{center}
\vspace{-0.5cm}
\caption{(Left) Results for $R_1+R_2$ on 10 ensembles from fits to $C_J^\text{3pt}$
  without including the $N \pi$ state, i.e., the spectrum taken
  from fits to $C^{\rm 2pt}$.  (Right) Including the $N \pi$
  state. For PCAC to be satisfied, $R_1 + R_2$  should be unity up to
  discretization errors. The dotted lines show the 5\% deviation band.\looseness-1}
\label{fig:deviation}
\end{figure}

The non-singlet PCAC relation between bare axial, $A_\mu(x)$, and
pseudoscalar, $P(x)$, currents is:\looseness-1
\begin{equation}
  \partial_\mu A_\mu = 2\widehat{m} P \,,
  \label{eq:PCAC-op}
\end{equation}
where the quark mass parameter $\widehat{m} \equiv Z_m m_{ud} Z_P
Z_A^{-1}$ includes all the renormalization factors, and $m_{ud} = (m_u
+ m_d)/2 =m_l$ is the light quark mass in the isospin symmetric limit.
Using the decomposition in Eqs.~\eqref{eq:aff-a} and~\eqref{eq:aff-ps}
of GSME, the PCAC relation requires that the three form factors $G_A$,
$\GP$, and $G_P$ must satisfy, up to discretization
errors, the relation
\begin{equation}
  2M_N G_A(Q^2) - \frac{Q^2}{2M_N}\GP(Q^2) = 2\widehat{m} G_P(Q^2) \,, 
\label{eq:PCAC2}
\end{equation}
on each ensemble. All pre Ref.~\cite{Rajan:2017lxk} calculations did
not check this relation and missed observing that the data showed large
deviations.  Calculations subsequent to Ref.~\cite{Jang:2019vkm} that
include the lowest mass gap state, $N({\bf p}=1) \pi({\bf p}=-1)$, in
the analysis, obtain form factors that satisfy PCAC to within $\approx
10\%$ already at lattice spacing of $a \approx 0.9$~fm. (The ETMC
result is an exception as explained in
Ref.~\cite{Alexandrou:2023qbg}).  An illustration of the size of the
deviation from unity of $R_1 + R_2 \equiv \frac{2\widehat{m}
  G_P(Q^2)}{M_N G_A(Q^2)} + \frac{Q^2 \GP(Q^2)}{ 4M_N^2 G)_A(Q^2)}$,
without and with the lowest $N \pi$ state included is shown in
Fig.~\ref{fig:deviation} taken from Ref.~\cite{Jang:2023zts}.

To summarize, satisfying the PCAC relation in Eq.~\eqref{eq:PCAC2}
provides a strong and necessary constraint on the extraction of the
three axial form factors. $\chi$PT analysis by
B\"ar~\cite{Bar:2018akl,Bar:2018xyi} and data driven validation in
Ref.~\cite{Jang:2019vkm,RQCD:2019jai,Barca:2022uhi} show that the
lowest, $N({\bf p}=1) \pi({\bf p}=-1)$ and $N({\bf p}=0) \pi({\bf p}=1)$,
states makes a large contribution on the two sides, respectively, and need
to be included in the analysis. For percent
level precision, the next question is---what other states need to be
included?  Current analyses include up to three states, where the
third state, if its parameters are left free, effectively tries to
account for all residual ESC.  Such fits have been implemented in
different ways. For example, in Ref.~\cite{RQCD:2019jai}, the $N \pi$
state is hardwired and the third state is taken to be the lowest
excited state in fits to $C^{\rm 2pt}$.  In
Refs.~\cite{Jang:2019vkm,Park:2021ypf,Jang:2023zts}, a simultaneous
fit to all five $J=A_\mu$ and $P$ correlators is made wherein the
$A_4$ correlator fixes $E_1$ to approximately the non-interacting
energy of the $N \pi$ state. Over time, with much higher statistics
data, results from different collaborations using different  methods should
converge as more more excited states are included.

\begin{figure*} [ht]     %F07
    \centering
    \includegraphics[width=0.57\textwidth]{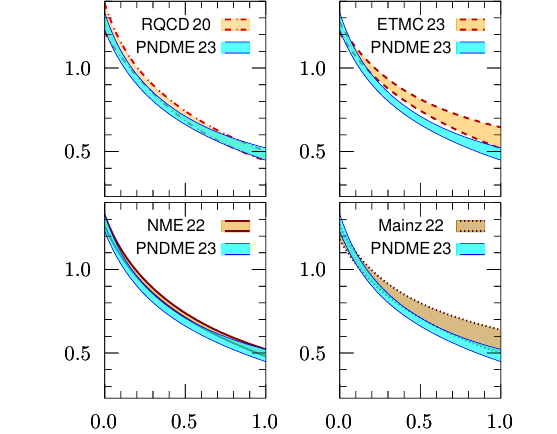}
    \hspace{0.cm}
    \includegraphics[width=0.41\textwidth]{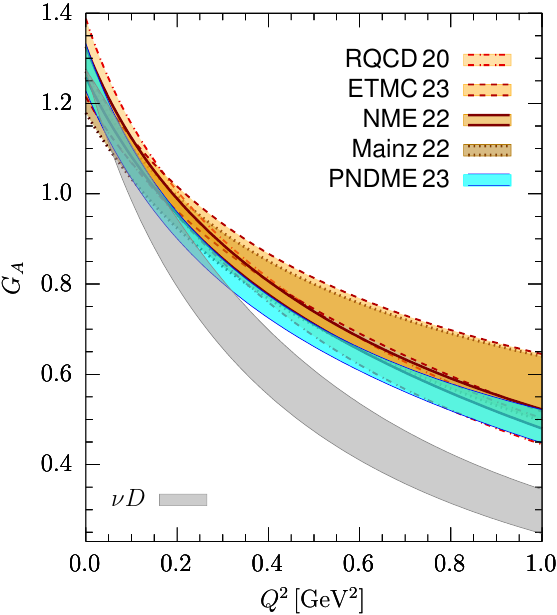}
    \caption{(Left) Comparison of the nucleon axial-vector form factor
      $G_A \left( Q^2 \right)$ as a function of $Q^2$, the momentum
      transfer squared, obtained by the
      PNDME 23~\protect\cite{Jang:2023zts} shown by the turquoise band;
      RQCD 19~\protect\cite{RQCD:2019jai} (light faun band);
      ETMC  21~\protect\cite{Alexandrou:2023qbg} (faun band);
      NME 22~\protect\cite{Park:2021ypf} (light brown band);
      and Mainz 22~\protect\cite{Djukanovic:2022wru} (brown band).
      The $\nu$D band is the fit to the old neutrino-deuterium data
      taken from Ref.~\protect\cite{Meyer:2016oeg}.}
\label{fig:AFF_lattice}
\end{figure*}

\subsection{Extrapolating lattice AVFF to the physical point for use in phenomenology}    %S02E
\label{sec:CCFV}

The next step, once ESC have been removed and form factors have been
extracted from GSME on each ensemble, is to extrapolate these data
to the physical point and provide a parameterized form for $G_A$ and
$\GP$ that can be used in phenomenology. The challenge is that the
discrete set of $Q^2_i$ values at which data are obtained are
different on each ensemble.

One, simple to implement, way consists of the following three steps:
\begin{enumerate}
\item Parameterize the $G_A(Q^2_i)$ data on each ensemble. Depending
  on the number of $Q^2_i$ values, it could be a suitably truncated
  $z$-expansion or a Pad\'e.  The expected $1/Q^4$ asymptotic behavior
  can be built in by using sum rules in the
  $z$-expansion~\cite{Lee:2015jqa} or through an $\{n,n+2\}$ Pad\'e in
  $Q^2/M_N^2$.  The Mainz collaboration~\cite{Djukanovic:2022wru}
  combines the removal of ESC at various values of $Q^2_i$ and the
  $Q^2$ parameterization on a given ensemble to include correlations.
\item Pick $n$ values of momenta, $Q_k^2$, over a range, say $ 0 \le Q^2 \le
  1$~GeV${}^2$.  Extrapolate the data at each of these values of
  $Q^2_k$ using a simultaneous fit in $\{m_q, a, M_\pi L$\} to the
  physical point. A typical ansatz used for such 
  chiral-continuum-finite-volume (CCFV) extrapolations is 
\begin{align}
g(M_\pi, a, M_\pi L) &= g_0 + c_1 a + c_2 M_\pi^2 
                       + \frac{c_3 M_\pi^2 \exp{(-M_\pi L)}}{\sqrt{M_\pi L}} \,,
  \label{eq:ccfv}
\end{align}
where I have kept only the lowest order corrections in each of the $\{m_q,
a,M_\pi L\}$ variables and assumed that discretization errors start at $O(a)$.
\item
Having obtained the form factor in the continuum limit at the $n$ 
points, $Q^2_k$, carry out the final parameterization,
using again a truncated $z$-expansion or a Pad\'e.
\end{enumerate}
This 3-step process can be done within a single bootstrap procedure to
propagate errors as has been done in
Ref.~\cite{Park:2021ypf,Jang:2023zts} to produce the NME and PNDME
results shown in Fig.~\ref{fig:AFF_lattice}.  Or these steps can be
combined, especially if there are correlations between them. For
example, to account for correlations between the
coefficients of the CCFV fits at different values of $Q^2_k$ in step 2. 

The plots in Fig.~\ref{fig:AFF_lattice} provide two comparisons. In panels on 
the left, physical point results from the
RQCD~\cite{RQCD:2019jai,Bali:2023sdi}, ETMC~\cite{Alexandrou:2023qbg},
NME~\cite{Park:2021ypf}, and Mainz~\cite{Djukanovic:2022wru}
collaborations are compared against those from
PNDME~\cite{Jang:2023zts}. On the right they are overlaid and compared
to the phenomenological extraction from the old neutrino-deuterium
bubble chamber data~\cite{Meyer:2016oeg}.  The PNDME, RQCD and NME
data mostly overlap, whereas the ETMC and Mainz data overlap and fall
off slightly slower for $Q^2 \gtrsim 0.3$~GeV${}^2$. On the other hand, the
neutrino-deuterium ($\nu$D) data~\cite{Meyer:2016oeg} falls off much
faster for $Q^2 \gtrsim 0.2$~GeV${}^2$. Overall, as shown in the right
plot, the five lattice QCD estimates are consistent within $1\sigma$
and lie about $2\sigma$ above the $\nu$D band for $Q^2 \gtrsim 0.3$~GeV${}^2$.

There also are results from CalLAT~\cite{Meyer:2021vfq,Meyer:2022mix},
PACS~\cite{Tsuji:2022ric,Tsuji:2023llh}, and
LHP+RBC+UKQCD~\cite{Ohta:2023ygq} collaborations, which have not been
included in the comparison because they have not been extrapolated to
the physical point.  The Fermilab collaboration~\cite{Grebe:2023tfx}
has embarked on the much harder problem of calculating transition
matrix elements as well, e.g., $N \to \Delta$ or $N \to N \pi$.

\begin{figure}[!tbh]  %F08
  \centering
  \includegraphics[width=0.48\textwidth]{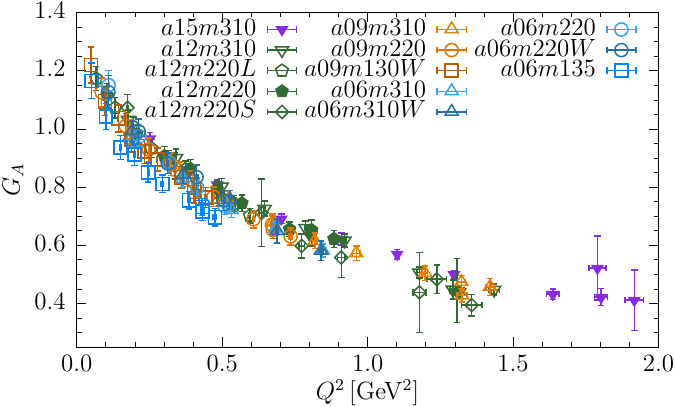} 
  \caption{The data for renormalized $G_A(Q^2)$ on 13 HISQ ensembles
    show small variation in $\{a,M_\pi,M_\pi L\}$. The $a06m135$ data
    are statistics limited. Figure reproduced from
    Ref.~\protect\cite{Jang:2023zts}.  }
  \label{fig:GAdata13}
\end{figure}

From the  analysis of the NME and PNDME data, my understanding is that the
differences in exactly how the ESC are handled by the various
collaborations and the consequent uncertainty in the final results should be
considered work in progress. The uncertainty from the differences in
the overall procedure for parameterization and CCFV extrapolation is,
I believe, smaller because the data do not show large
dependence on any of the three parameters $\{m_q, a,M_\pi L\}$, especially for 
$a \lesssim 0.1$~fm and $M_\pi L\gtrsim 4$, as illustrated in
Fig.~\ref{fig:GAdata13}~\cite{Park:2021ypf,Jang:2023zts}. Hopefully,
the next generation calculations will shed light on, and possibly
resolve, the various differences.

Other findings in Refs.~\cite{Park:2021ypf,Jang:2023zts} are (i) the
dipole ansatz $G_A(Q^2) = \frac{g_A}{(1 + c Q^2/M_N^2)^2}$ gives poor
fits (very low $p$ values) to data on many ensembles. Our conclusion,
therefore, is that the lattice data already show that the dipole
ansatz does not have enough parameters to capture the $Q^2$ behavior
over the range $ 0 \le Q^2 \le 1$~GeV${}^2$.
(ii) The PPD relation between $G_A$ and $\GP$ works very well. \looseness-1

\begin{figure*}     %F09
    \centering
    \includegraphics[width=0.30\textwidth]{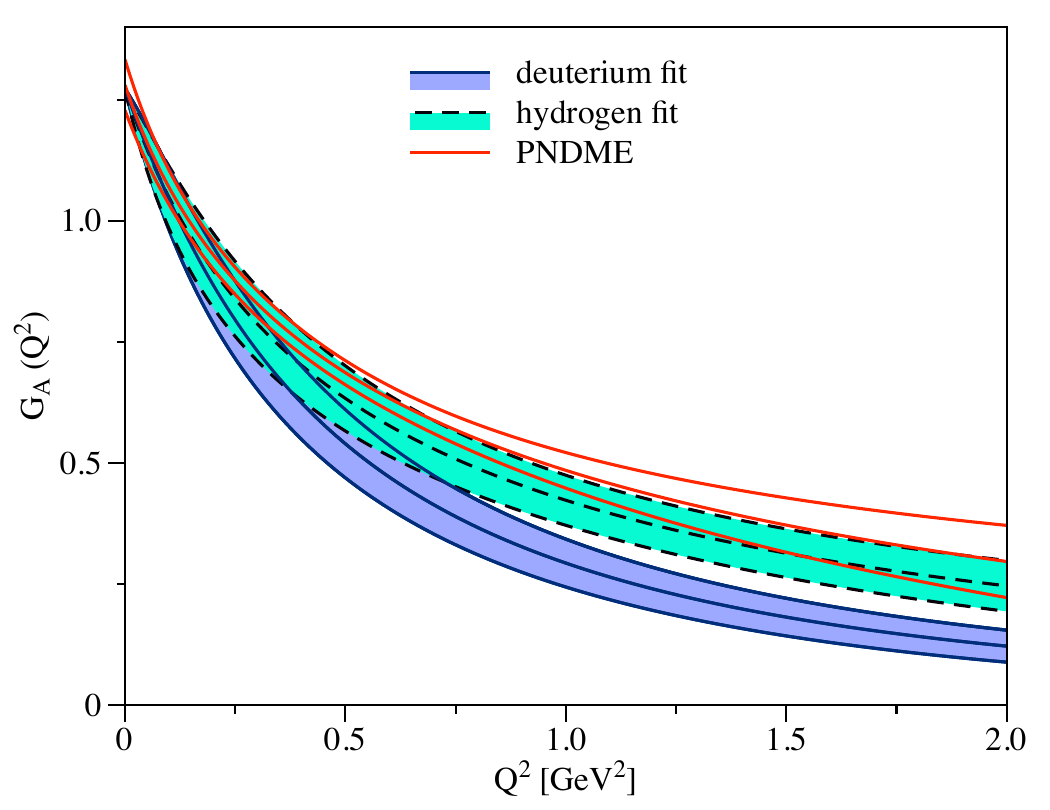}
    \hspace{0.cm}
    \includegraphics[width=0.68\textwidth]{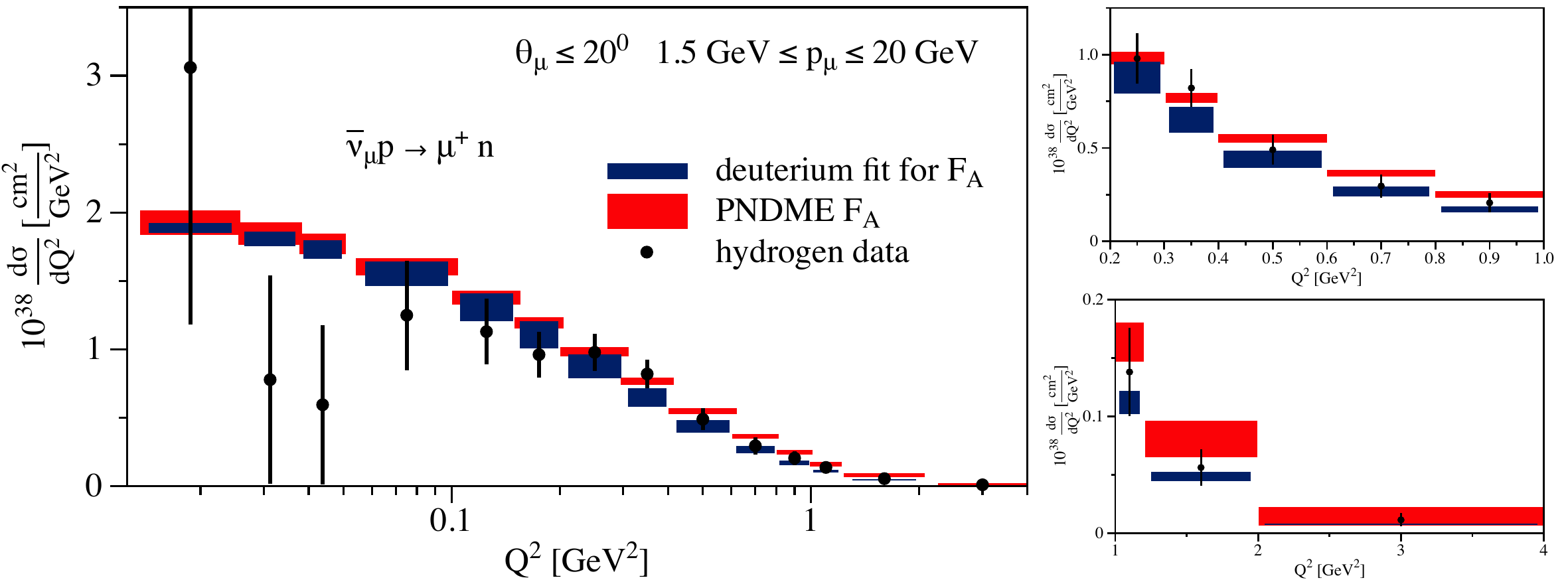}
    \caption{(Left) Comparison of the parameterized nucleon axial-vector form factor
      $G_A ( Q^2)$ versus  $Q^2$ up to $2$~GeV${}^2$ obtained from
      (i) fit to the deuterium bubble-chamber
      data~\protect\cite{Meyer:2016oeg} shown by blue solid lines with
      error band; (ii) fit to recent MINERvA antineutrino-hydrogen
      data~\protect\cite{MINERvA:2023avz}, shown by black dashed lines
      and turquoise error band; and (iii) lattice QCD result obtained
      by the PNDME Collaboration~\protect\cite{Jang:2023zts} shown by
      red solid lines {\it{without a band}}. (Right) A comparison of
      antineutrino-nucleon charged current elastic differential cross
      section using AFF from (i) lattice QCD by the PNDME
      collaboration~\protect\cite{Jang:2023zts} (red bands) and (ii)
      the deuterium bubble-chamber data~\protect\cite{Meyer:2016oeg}
      (black bands) with the MINERvA antineutrino-hydrogen
      data~\protect\cite{MINERvA:2023avz} (black circles). These
      figures are taken from Ref.~\protect\cite{Tomalak:2023pdi}.}
\label{fig:Minerva}
\end{figure*}

\subsection{Consistency check in the extraction of the axial charge $g_A^{u-d}$}     %S02F
\label{sec:Acharge}

There are two ways in which one can extract the axial charge $g_A^{u-d}$.  The
first is from the forward matrix element using $C_{A_3}$
in Eq.~\eqref{eq:A3-decomp} with ${\bm q} = 0$ and the second is by
extrapolating the form factor $G_A(Q^2 \neq 0)$ to $Q^2 = 0$. I am
considering them as separate because the extraction from the forward
matrix element is computationally clean: $C_{A_3}({\bm q} = 0)$ has the smallest errors 
and verification of the symmetry of the data about $\tau/2$ is a good test. The errors
grow with ${\bm q}$ as shown in Fig.~\ref{fig:ESCcomp}. On the other
hand, $G_A(Q^2)$ is constrained by being part of the PCAC 
relation, Eq.~\eqref{eq:PCAC2}, that has to be satisfied. The two
results must agree after CCFV extrapolation. Based on the data in
Ref.~\cite{Jang:2023zts}, I conclude\looseness-1
\begin{itemize}
\item
The difference between $g_A^{u-d}$ extracted without and with
including $N \pi$ states is $O(\approx 6\%)$, i.e., $1.218(39) \to
1.294(48)$ on including one (the lowest) $N\pi$ state in the analysis. Note that
the errors in each result are $O(\approx 3\%)$.
\item
The difference between $g_A^{u-d} \equiv G_A(Q^2 \to 0)$ extracted by
extrapolating $G_A(Q^2)$ data obtained without and with including the
lowest $N \pi$ state is also $O(\approx 6\%)$, i.e., $1.213(39) \to
1.289(56)$.  Again, the errors in each are $O(\approx 4\%)$.
\end{itemize}
Thus, for each of the two cases, without and with $N \pi$
states, we get consistent estimates for  $g_A^{u-d}$ from the two
methods, however results including the $N \pi$ state are about $6\%$
larger.  This difference is consistent with the  expected $\sim 5\%$ 1-loop 
correction to the charge in $\chi$PT, however, it 
is roughly one combined $\sigma$. It therefore needs validation. 
My pick for the final result is the analysis including the $N
\pi$ state since it gives form factors that satisfy the PCAC relation. 
Higher precision data are needed to further clarify 
the other significant ESCs and how to include them.

\begin{table*}[htb]   %T01
  % full CCFV fits
  \centering
  \renewcommand{\arraystretch}{1.1}
%  \begin{ruledtabular}
    \begin{tabular}{ccccc}
Collaboration & $g_A^{u-d}$   & $\expv{r_A^2}$~fm${}^2$ & $g_P^\ast$   & $g_{\pi NN}$ \\ \hline
PNDME 23  & 1.292(53)(24)     &  0.439(56)(34)  & 9.03(47)(42)  &  14.14(81)(85) \\
RQCD 19/23& $1.284^{28}_{27}$ & 0.449(88)       &  8.68(45) & 12.93(80) \\
ETMC 23   & 1.283(22)         &  0.339(46)(6)   & 8.99(39)(49)    &  13.25(67)(69)\\
PACS 23   & 1.264(14)(1)      &  0.316(67)      &    &  \\
Mainz 22  & 1.225(39)(25)     &  0.370(63)(16)  &   &  \\
NME  21   & 1.32(6)(5)        &  0.428(53)(30)  &  7.9(7)(9)  & 12.4(1.2) \\
\hline
CalLat 18  & 1.271(10)(7)     &                 &             &     \\
PNDME 18   & 1.218(25)(30)    &                 &             &     \\
Mainz 19   & 1.242(25)(${}^{0}_{-30})$   &                 &             &     \\
$\chi$QCD 18 & 1.254(16)(30)    &                 &             &     \\
\hline
    \end{tabular}
%  \end{ruledtabular}
  \caption{Comparison of $g_{A}$, $\expv{r_A^2}$, $g_P^\ast$ and
    $g_{\pi NN}$ from recent calculations: 
    PNDME 23~\protect\cite{Jang:2023zts},  
    RQCD 19/23~\protect\cite{RQCD:2019jai,Bali:2023sdi},
    ETMC 23~\protect\cite{Alexandrou:2020okk},
    PACS 23~\protect\cite{Tsuji:2023llh,Tsuji:2022ric}, 
    Mainz 22~\protect\cite{Djukanovic:2022wru}, and
    NME 21~\protect\cite{Park:2021ypf}.
    Lattice results for all charges are usually quoted 
    in the $\overline{\rm MS}$ scheme at scale 2~GeV, however, note that 
    the renormalization of $g_A^{u-d}$ is trivial~\protect\cite{FlavourLatticeAveragingGroupFLAG:2021npn}. Earlier
    results for $g_A^{u-d}$ with 2+1+1-flavor simulations by 
    CalLat 18~\cite{Chang:2018uxx} and 
    PNDME 18~\cite{Gupta:2018qil}, and with 2+1-flavor simulations by 
    Mainz 19~\cite{Harris:2019bih} and 
    $\chi$QCD 18~\cite{Liang:2018pis} 
    that entered in the averages compiled in the
    FLAG report 2021~\protect\cite{FlavourLatticeAveragingGroupFLAG:2021npn} are given below the dividing line. }
  \label{tab:final_Comp}
\end{table*}

\section{Comparison of charges obtained by various lattice collaborations}      %S03
\label{sec:charges}

The results for the axial charge, $g_A^{u-d}$, the charge radius
squared, $\expv{r_A^2}$, the induced pseudoscalar coupling $g_P^\ast$,
and the pion-nucleon coupling $g_{\pi NN}$ , extracted from $G_A$ and $\GP$ 
by various collaborations using the relations 
\begin{align}
  G_A(Q^2) &= g_A ( 1 - \frac{\expv{r_A^2}}{6} Q^2 + \cdots) \,, \\
  g_P^\ast &\equiv \frac{m_\mu}{2M_N}\GP(Q^{\ast 2}) \,, \\
  g_{\pi NN} &\equiv \;  \lim_{Q^2 \to -M_\pi^2} \frac{M_\pi^2 + Q^2}{4M_N F_\pi} {\widetilde G}_P(Q^2) \,.
  \label{eq:gPstar}
\end{align}
are summarized in
Table~\ref{tab:final_Comp}.  Here $m_\mu$ is the muon mass and
$Q^{\ast 2} = 0.88 m_\mu^2$ is the energy scale of muon capture, and
$F_\pi=92.9$~MeV is the pion decay constant.

The results show about $10\%$ variation in $g_A^{u-d}$, $g_P^\ast$,
and $g_{\pi NN}$ and about $25\%$ in $\expv{r_A^2}$. Part of this is
likely due to different methodologies used in the analysis, in particular how and if
the lowest $N\pi$ state is included in the analysis. These results will 
improve steadily over time with higher statistics data. 

\section{Comparison of differential cross-section using lattice AVFF with MINER$\nu$A data\looseness-1}      %S04
\label{sec:Minerva}

A comparison of the antineutrino-nucleon charged-current elastic cross
sections calculated using predictions of AVFF from lattice (PNDME
23~\cite{Jang:2023zts}) and neutrino-deuterium
analysis~\cite{Meyer:2016oeg} with MINERvA
measurement~\cite{MINERvA:2023avz,Irani:2023lol} is presented in
Fig.~\ref{fig:Minerva} reproduced from Ref.~\cite{Tomalak:2023pdi}.  A
\(\chi^2\) test was performed to determine the significance of the
differences between the three.  No significant difference was found
between MINERvA-lattice QCD (PNDME) and between MINERvA-deuterium
results.  A $\approx 2.5\sigma$ tension was, however, found between
the PNDME and the deuterium results.  Based on data shown in
Fig.~\ref{fig:AFF_lattice}, the deviation in the deuterium-Mainz and
deuterium-ETMC will be even larger. To assess the scope for future
progress, three regions of $Q^2$ with different prospects for the
extraction of AVFF from lattice QCD and MINERvA-like experiments were
identified.

For $Q^2 \lesssim 0.2~\mathrm{GeV}^2$, LQCD predictions and fits to
the deuterium bubble-chamber data are in good agreement. In this
region, the experimental errors in the measurement on hydrogen by
MINERvA are large, whereas the errors in the parameterization of the
deuterium bubble chamber data are smaller. The $\nu$D result has
often been used as a benchmark, however, note that there is
unresolved uncertainty in the deuterium data as discussed in
Ref.~\cite{Meyer:2016oeg}. Also, no new deuterium data are expected in
the near-term, so I do not comment on its future prospects. Lattice
QCD data are competitive and will improve steadily. This region will be
well-characterized by the axial charge, the axial charge radius, and
well-parameterized by a low-order $z$-expansion or a Pad\'e.

For $0.2~\mathrm{GeV}^2 \lesssim Q^2 \lesssim 1~\mathrm{GeV}^2$, the
AVFF from PNDME has the smallest errors and the predicted
differential cross section lies above the hydrogen and $\nu$D
values, i.e., the same ordering as for the AVFF shown in
Fig.~\ref{fig:Minerva} (left). Future improvements in both the
hydrogen data and lattice calculations will provide robust
cross-checks in this region.

The region $Q^2 \lesssim 0.5~\mathrm{GeV}^2$ is where lattice QCD
data, even with current methodology, will improve rapidly as more
simulations are done closer to $M_\pi=135$~MeV, $a \to 0$ and on
larger volumes because, in these limits and for given statistics, the
value of $Q^2|_{\rm max}$ with a good signal (usually taken to be some
fixed lattice momentum $\textbf n^2$) decreases.

For $Q^2 \gtrsim 1~\mathrm{GeV}^2$, current LQCD data have larger
statistical errors and systematic uncertainties---discretization and
residual excited state contributions. With the current methodology,
the lattice AVFF comes mostly from simulations with $M_\pi \gtrsim
300$~MeV on $a < 0.1$~fm ensembles~\cite{Jang:2023zts}. If the
dependence on $\{a,M_\pi\}$ is mild, as has been observed so far, then
these data are useful by themselves.  With higher statistics and
improved actions, one can push the lattice momenta ${\bm n}^2|_{\rm
  max}$ higher and perhaps reliably reach $Q^2 \sim
2\ \mathrm{GeV}^2$. Nevertheless, new methods are definitely needed to
get data at $Q^2 \gtrsim 2\ \mathrm{GeV}^2$ from simulations with
physical pion masses, $M_\pi \approx 135$~MeV and $a <
0.1$~fm. Similarly, improvements in MINERvA and follow on experiments
are needed to cover the full range of $Q^2$ relevant for DUNE.

\section{Concluding remarks}      %S05
\label{sec:conclusions}

Extensive calculations of the AVFF are being carried out by at least the following nine 
lattice QCD collaborations: PNDME~\cite{Jang:2023zts}, 
RQCD~\cite{RQCD:2019jai,Bali:2023sdi}, 
ETMC~\cite{Alexandrou:2023qbg}, 
NME~\cite{Park:2021ypf}, 
Mainz~\cite{Djukanovic:2022wru}, 
CalLAT~\cite{Meyer:2021vfq,Meyer:2022mix}, 
PACS~\cite{Tsuji:2022ric,Tsuji:2023llh}, 
LHP+RBC+UKQCD~\cite{Ohta:2023ygq}, and 
Fermilab~\cite{Grebe:2023tfx}. As shown in Sections~\ref{sec:charges} and~\ref{sec:CCFV}, 
we now have results to within 10\% precision.  The major uncertainty comes from 
resolving and removing excited-state contributions. 

The good news is that the methodology for the calculation of the
correlation functions, $C^{\rm 2pt}$ and $C_J^{\rm 3pt}$, is
robust. The bad news is that the exponentially falling signal to noise
ratio in them means that ESC are large at source-sink separations
possible in today's calculations. Second, it is also clear that
multihadron, $N \pi$, excited states give large contributions which must
be removed. Unfortunately, it is not
yet known how many of these states need to be included in the
analysis for percent level precision. The operator constraint that the
form factors satisfy the PCAC relation in Eq.~\eqref{eq:PCAC2}
provides a valuable check, so it must be carried out in all
calculations. The third challenge is getting data at large $Q^2$
because the discretization and statistical errors grow with $Q^2$ on a
given ensemble. Also, the $Q^2|_{\rm max}$ (the largest lattice momenta
${2\pi {\bm n}}/{La}$ with a good signal to noise ratio for fixed
statistics) decreases as simulations are done closer to the physical
point. Thus, to get data for $Q^2 \gtrsim 1$~GeV${}^2$ on physical
pion mass ensembles will need/benefit from new methods and very
high statistics.\looseness-1

As first step towards percent level precision, my estimate is that  a
factor of ten increase in statistics will reduce the statistical
errors to a level that will provide much more clarity in removing
ESC. Similary new developments, including variational
methods~\cite{Barca:2022uhi} with multihadron states and momentum
smearing~\cite{Bali:2016lva}, will improve the calculations and extend
the range of $Q^2$. I anticipate continued improvements in both,
statistics and methods, will provide LQCD predictions of AVFF for
nucleons in the range $Q^2 \lesssim 2~\mathrm{GeV}^2$ (hopefully
higher) with percent level precision by about 2030, in concert with
DUNE producing data.\looseness-1

\acknowledgments

Many thanks to my collaborators, Yong-Chull Jang, Sungwoo Park,
Oleksandr Tomalak, Tanmoy Bhattacharya, Vincenzo Cirigliano, Huey-Wen
Lin, Emanuele Mereghetti, Santanu Mondal, and Boram Yoon for making
the PNDME and NME results happen. Also to all who presented relevant results 
at Lattice 2023 and to Andr\'e Walker-Laud for valuable comments. Our calculations used the Chroma
software suite~\cite{Edwards:2004sx} and resources at (i) the National
Energy Research Scientific Computing Center, a DOE Office of Science
User Facility supported by the Office of Science of the
U.S. Department of Energy under Contract No. DE-AC02-05CH11231; (ii)
the Oak Ridge Leadership Computing Facility, which is a DOE Office of
Science User Facility supported under Contract DE-AC05-00OR22725,
through awards under the ALCC program project LGT107 and INCITE award
HEP133; (iii) the USQCD collaboration, which is funded by the Office
of Science of the U.S. Department of Energy; and (iv) Institutional
Computing at Los Alamos National Laboratory.  R.~Gupta was partly
supported by the U.S. Department of Energy, Office of Science, Office
of High Energy Physics under Contract No. DE-AC52-06NA25396 and by
the LANL LDRD program.

%%%%%%%%%%%%%%%%%%%%%%%%%%%%%%%%%%%%%%%%%%%%%%%%%%%%%%%%%%%%%%%%%%%%%%%
\bibliographystyle{JHEP}
\let\oldbibitem\bibitem
\def\bibitem#1\emph#2,{\oldbibitem#1}
\let\oldthebibliography\thebibliography
\renewcommand\thebibliography[1]{\oldthebibliography{#1}%
                                 \itemsep0pt\parskip0pt\relax}
\bibliography{refs}

% \bibliographystyle{JHEP}
%\bibliographystyle{apsrev}
%\let\origbibitem\bibitem
%\def\bibitem#1#2\emph#3, {\origbibitem{#1}#2}
%\bibliography{ref} %%% ref.bib file

%% \begin{thebibliography}{99}
%% \bibitem{...}
%% ....
%% \end{thebibliography}

\end{document}

%% file: macro.tex
\providecommand{\wbar}[1]{\overline#1}
\providecommand{\abs}[1]{\lvert#1\rvert}

\providecommand{\bra}[1]{\langle#1\rvert}
\providecommand{\ket}[1]{\lvert#1\rangle}
\providecommand{\matrixe}[3]{\langle#1\lvert#2\rvert#3\rangle}
\providecommand{\expv}[1]{\langle#1\rangle}

\newcommand{\Dslash}{\ensuremath{{D\kern -0.65em /}}}
\renewcommand{\Re}{\mathop{\mathrm{Re}}}
\renewcommand{\Im}{\mathop{\mathrm{Im}}}

\newcommand{\GP}{\widetilde{G}_P}

\definecolor{green}{rgb}{0.1, 0.8, 0.1}

%% file: main.bbl
\providecommand{\href}[2]{#2}\begingroup\raggedright\begin{thebibliography}{10}

\bibitem{Mendenhall:2012tz}
{\scshape UCNA Collaboration} collaboration, M.~Mendenhall et~al.,
  \emph{{Precision measurement of the neutron $\beta$-decay asymmetry}},
  \href{https://doi.org/10.1103/PhysRevC.87.032501}{\emph{Phys.Rev.} {\bfseries
  C87} (2013) 032501} [\href{https://arxiv.org/abs/1210.7048}{{\ttfamily
  1210.7048}}].

\bibitem{Brown:2017mhw}
{\scshape UCNA} collaboration, M.~A.~P. Brown et~al., \emph{{New result for the
  neutron $\beta$-asymmetry parameter $A_0$ from UCNA}},
  \href{https://doi.org/10.1103/PhysRevC.97.035505}{\emph{Phys. Rev.}
  {\bfseries C97} (2018) 035505}
  [\href{https://arxiv.org/abs/1712.00884}{{\ttfamily 1712.00884}}].

\bibitem{Markisch:2018ndu}
B.~M\"arkisch et~al., \emph{{Measurement of the Weak Axial-Vector Coupling
  Constant in the Decay of Free Neutrons Using a Pulsed Cold Neutron Beam}},
  \href{https://doi.org/10.1103/PhysRevLett.122.242501}{\emph{Phys. Rev. Lett.}
  {\bfseries 122} (2019) 242501}
  [\href{https://arxiv.org/abs/1812.04666}{{\ttfamily 1812.04666}}].

\bibitem{Mund:2012fq}
D.~Mund, B.~Maerkisch, M.~Deissenroth, J.~Krempel, M.~Schumann, H.~Abele
  et~al., \emph{{Determination of the Weak Axial Vector Coupling from a
  Measurement of the Beta-Asymmetry Parameter A in Neutron Beta Decay}},
  \href{https://doi.org/10.1103/PhysRevLett.110.172502}{\emph{Phys. Rev. Lett.}
  {\bfseries 110} (2013) 172502}
  [\href{https://arxiv.org/abs/1204.0013}{{\ttfamily 1204.0013}}].

\bibitem{Ademollo:1964sr}
M.~Ademollo and R.~Gatto, \emph{{Nonrenormalization Theorem for the Strangeness
  Violating Vector Currents}},
  \href{https://doi.org/10.1103/PhysRevLett.13.264}{\emph{Phys.Rev.Lett.}
  {\bfseries 13} (1964) 264}.

\bibitem{Donoghue:1990ti}
J.~F. Donoghue and D.~Wyler, \emph{{Isospin breaking and the precise
  determination of $V_{ud}$}},
  \href{https://doi.org/10.1016/0370-2693(90)91287-L}{\emph{Phys.Lett.}
  {\bfseries B241} (1990) 243}.

\bibitem{Bhattacharya:2011qm}
T.~Bhattacharya, V.~Cirigliano, S.~D. Cohen, A.~Filipuzzi, M.~Gonzalez-Alonso,
  M.~L. Graesser et~al., \emph{{Probing Novel Scalar and Tensor Interactions
  from (Ultra)Cold Neutrons to the LHC}},
  \href{https://doi.org/10.1103/PhysRevD.85.054512}{\emph{Phys. Rev. D}
  {\bfseries 85} (2012) 054512}
  [\href{https://arxiv.org/abs/1110.6448}{{\ttfamily 1110.6448}}].

\bibitem{Ivanov:2014bya}
A.~N. Ivanov, M.~Pitschmann, N.~I. Troitskaya and Y.~A. Berdnikov,
  \emph{{Bound-state $\beta$\ensuremath{-}decay of the neutron re-examined}},
  \href{https://doi.org/10.1103/PhysRevC.89.055502}{\emph{Phys. Rev. C}
  {\bfseries 89} (2014) 055502}
  [\href{https://arxiv.org/abs/1401.7809}{{\ttfamily 1401.7809}}].

\bibitem{Czarnecki:2018okw}
A.~Czarnecki, W.~J. Marciano and A.~Sirlin, \emph{Neutron lifetime and axial
  coupling connection},
  \href{https://doi.org/10.1103/PhysRevLett.120.202002}{\emph{Phys.\ Rev.\
  Lett.} {\bfseries 120} (2018) 202002}.

\bibitem{Czarnecki:2019mwq}
A.~Czarnecki, W.~J. Marciano and A.~Sirlin, \emph{{Radiative Corrections to
  Neutron and Nuclear Beta Decays Revisited}},
  \href{https://doi.org/10.1103/PhysRevD.100.073008}{\emph{Phys.\ Rev.}
  {\bfseries D100} (2019) 073008}
  [\href{https://arxiv.org/abs/1907.06737}{{\ttfamily 1907.06737}}].

\bibitem{Czarnecki:2019iwz}
A.~Czarnecki, W.~J. Marciano and A.~Sirlin, \emph{{Pion beta decay and
  Cabibbo-Kobayashi-Maskawa unitarity}},
  \href{https://doi.org/10.1103/PhysRevD.101.091301}{\emph{Phys.\ Rev.}
  {\bfseries D101} (2020) 091301}
  [\href{https://arxiv.org/abs/1911.04685}{{\ttfamily 1911.04685}}].

\bibitem{Horoi:2018fls}
M.~Horoi and A.~Neacsu, \emph{{Shell model study of using an effective field
  theory for disentangling several contributions to neutrinoless double-$\beta$
  decay}}, \href{https://doi.org/10.1103/PhysRevC.98.035502}{\emph{Phys. Rev.
  C} {\bfseries 98} (2018) 035502}
  [\href{https://arxiv.org/abs/1801.04496}{{\ttfamily 1801.04496}}].

\bibitem{Carroll2007}
B.~W. Carroll and D.~A. Ostlie, \emph{{A}n {I}ntroduction to {M}odern
  {A}strophysics}. 2nd (international)~ed., 2007.

\bibitem{Ruso:2022qes}
L.~Alvarez~Ruso et~al., \emph{{Theoretical tools for neutrino scattering:
  interplay between lattice QCD, EFTs, nuclear physics, phenomenology, and
  neutrino event generators}},
  \href{https://arxiv.org/abs/2203.09030}{{\ttfamily 2203.09030}}.

\bibitem{Kronfeld:2019nfb}
{\scshape USQCD} collaboration, A.~S. Kronfeld, D.~G. Richards, W.~Detmold,
  R.~Gupta, H.-W. Lin, K.-F. Liu et~al., \emph{{Lattice QCD and
  Neutrino-Nucleus Scattering}},
  \href{https://doi.org/10.1140/epja/i2019-12916-x}{\emph{Eur. Phys. J.}
  {\bfseries A55} (2019) 196}
  [\href{https://arxiv.org/abs/1904.09931}{{\ttfamily 1904.09931}}].

\bibitem{Meyer:2022mix}
A.~S. Meyer, A.~Walker-Loud and C.~Wilkinson, \emph{{Status of Lattice QCD
  Determination of Nucleon Form Factors and their Relevance for the Few-GeV
  Neutrino Program}},
  \href{https://doi.org/10.1146/annurev-nucl-010622-120608}{\emph{Ann. Rev.
  Nucl. Part. Sci.} {\bfseries 72} (2022) 205}
  [\href{https://arxiv.org/abs/2201.01839}{{\ttfamily 2201.01839}}].

\bibitem{MINERvA:2023avz}
{\scshape MINERvA} collaboration, T.~Cai et~al., \emph{{Measurement of the
  axial vector form factor from antineutrino\textendash{}proton scattering}},
  \href{https://doi.org/10.1038/s41586-022-05478-3}{\emph{Nature} {\bfseries
  614} (2023) 48}.

\bibitem{Tomalak:2023pdi}
O.~Tomalak, R.~Gupta and T.~Bhattacharya, \emph{{Confronting the axial-vector
  form factor from lattice QCD with MINERvA antineutrino-proton data}},
  \href{https://doi.org/10.1103/PhysRevD.108.074514}{\emph{Phys. Rev. D}
  {\bfseries 108} (2023) 074514}
  [\href{https://arxiv.org/abs/2307.14920}{{\ttfamily 2307.14920}}].

\bibitem{Jang:2023zts}
{\scshape Precision Neutron Decay Matrix Elements (PNDME)} collaboration, Y.-C.
  Jang, R.~Gupta, T.~Bhattacharya, B.~Yoon and H.-W. Lin, \emph{{Nucleon
  isovector axial form factors}},
  \href{https://doi.org/10.1103/PhysRevD.109.014503}{\emph{Phys. Rev. D}
  {\bfseries 109} (2024) 014503}
  [\href{https://arxiv.org/abs/2305.11330}{{\ttfamily 2305.11330}}].

\bibitem{Meyer:2016oeg}
A.~S. Meyer, M.~Betancourt, R.~Gran and R.~J. Hill, \emph{{Deuterium target
  data for precision neutrino-nucleus cross sections}},
  \href{https://doi.org/10.1103/PhysRevD.93.113015}{\emph{Phys. Rev. D}
  {\bfseries 93} (2016) 113015}
  [\href{https://arxiv.org/abs/1603.03048}{{\ttfamily 1603.03048}}].

\bibitem{FlavourLatticeAveragingGroupFLAG:2021npn}
{\scshape Flavour Lattice Averaging Group (FLAG)} collaboration, Y.~Aoki
  et~al., \emph{{FLAG Review 2021}},
  \href{https://doi.org/10.1140/epjc/s10052-022-10536-1}{\emph{Eur. Phys. J. C}
  {\bfseries 82} (2022) 869}
  [\href{https://arxiv.org/abs/2111.09849}{{\ttfamily 2111.09849}}].

\bibitem{FlavourLatticeAveragingGroup:2019iem}
{\scshape Flavour Lattice Averaging Group} collaboration, S.~Aoki et~al.,
  \emph{{FLAG Review 2019: Flavour Lattice Averaging Group (FLAG)}},
  \href{https://doi.org/10.1140/epjc/s10052-019-7354-7}{\emph{Eur. Phys. J. C}
  {\bfseries 80} (2020) 113}
  [\href{https://arxiv.org/abs/1902.08191}{{\ttfamily 1902.08191}}].

\bibitem{Alexandrou:2023qbg}
C.~Alexandrou, S.~Bacchio, M.~Constantinou, J.~Finkenrath, R.~Frezzotti,
  B.~Kostrzewa et~al., \emph{{Nucleon axial and pseudoscalar form factors using
  twisted-mass fermion ensembles at the physical point}},
  \href{https://arxiv.org/abs/2309.05774}{{\ttfamily 2309.05774}}.

\bibitem{Bali:2023sdi}
G.~S. Bali, S.~Collins, S.~Heybrock, M.~L\"offler, R.~R\"odl, W.~S\"oldner
  et~al., \emph{{Octet baryon isovector charges from $N_f = 2 + 1$ lattice
  QCD}},  \href{https://arxiv.org/abs/2305.04717}{{\ttfamily 2305.04717}}.

\bibitem{Tsuji:2022ric}
{\scshape PACS} collaboration, R.~Tsuji, N.~Tsukamoto, Y.~Aoki, K.-I. Ishikawa,
  Y.~Kuramashi, S.~Sasaki et~al., \emph{{Nucleon isovector couplings in Nf=2+1
  lattice QCD at the physical point}},
  \href{https://doi.org/10.1103/PhysRevD.106.094505}{\emph{Phys. Rev. D}
  {\bfseries 106} (2022) 094505}
  [\href{https://arxiv.org/abs/2207.11914}{{\ttfamily 2207.11914}}].

\bibitem{Djukanovic:2022wru}
D.~Djukanovic, G.~von Hippel, J.~Koponen, H.~B. Meyer, K.~Ottnad, T.~Schulz
  et~al., \emph{{Isovector axial form factor of the nucleon from lattice QCD}},
  \href{https://doi.org/10.1103/PhysRevD.106.074503}{\emph{Phys. Rev. D}
  {\bfseries 106} (2022) 074503}
  [\href{https://arxiv.org/abs/2207.03440}{{\ttfamily 2207.03440}}].

\bibitem{Park:2021ypf}
{\scshape Nucleon Matrix Elements (NME)} collaboration, S.~Park, R.~Gupta,
  B.~Yoon, S.~Mondal, T.~Bhattacharya, Y.-C. Jang et~al., \emph{{Precision
  nucleon charges and form factors using (2+1)-flavor lattice QCD}},
  \href{https://doi.org/10.1103/PhysRevD.105.054505}{\emph{Phys. Rev. D}
  {\bfseries 105} (2022) 054505}
  [\href{https://arxiv.org/abs/2103.05599}{{\ttfamily 2103.05599}}].

\bibitem{Alexandrou:2020okk}
C.~Alexandrou et~al., \emph{{Nucleon axial and pseudoscalar form factors from
  lattice QCD at the physical point}},
  \href{https://doi.org/10.1103/PhysRevD.103.034509}{\emph{Phys. Rev. D}
  {\bfseries 103} (2021) 034509}
  [\href{https://arxiv.org/abs/2011.13342}{{\ttfamily 2011.13342}}].

\bibitem{RQCD:2019jai}
{\scshape RQCD} collaboration, G.~S. Bali, L.~Barca, S.~Collins, M.~Gruber,
  M.~L\"offler, A.~Sch\"afer et~al., \emph{{Nucleon axial structure from
  lattice QCD}}, \href{https://doi.org/10.1007/JHEP05(2020)126}{\emph{JHEP}
  {\bfseries 05} (2020) 126}
  [\href{https://arxiv.org/abs/1911.13150}{{\ttfamily 1911.13150}}].

\bibitem{Babich:2010qb}
R.~Babich, J.~Brannick, R.~C. Brower, M.~A. Clark, T.~A. Manteuffel, S.~F.
  McCormick et~al., \emph{{Adaptive multigrid algorithm for the lattice
  Wilson-Dirac operator}},
  \href{https://doi.org/10.1103/PhysRevLett.105.201602}{\emph{Phys. Rev. Lett.}
  {\bfseries 105} (2010) 201602}
  [\href{https://arxiv.org/abs/1005.3043}{{\ttfamily 1005.3043}}].

\bibitem{Barca:2022uhi}
L.~Barca, G.~Bali and S.~Collins, \emph{{Toward N to N\ensuremath{\pi} matrix
  elements from lattice QCD}},
  \href{https://doi.org/10.1103/PhysRevD.107.L051505}{\emph{Phys. Rev. D}
  {\bfseries 107} (2023) L051505}
  [\href{https://arxiv.org/abs/2211.12278}{{\ttfamily 2211.12278}}].

\bibitem{Grebe:2023tfx}
A.~V. Grebe and M.~Wagman, \emph{{Nucleon-Pion Spectroscopy from Sparsened
  Correlators}},  \href{https://arxiv.org/abs/2312.00321}{{\ttfamily
  2312.00321}}.

\bibitem{Gusken:1989ad}
S.~Gusken, U.~Low, K.~H. Mutter, R.~Sommer, A.~Patel and K.~Schilling,
  \emph{{Nonsinglet Axial Vector Couplings of the Baryon Octet in Lattice
  {QCD}}}, \href{https://doi.org/10.1016/S0370-2693(89)80034-6}{\emph{Phys.
  Lett. B} {\bfseries 227} (1989) 266}.

\bibitem{Parisi:1983ae}
G.~Parisi, \emph{{The Strategy for Computing the Hadronic Mass Spectrum}},
  \href{https://doi.org/10.1016/0370-1573(84)90081-4}{\emph{Phys. Rept.}
  {\bfseries 103} (1984) 203}.

\bibitem{Lepage:1989hd}
G.~P. Lepage, \emph{{The Analysis of Algorithms for Lattice Field Theory}},  in
  \emph{{Boulder ASI 1989:97-120}}, pp.~97--120, 1989,
  \href{http://alice.cern.ch/format/showfull?sysnb=0117836}{http://alice.cern.ch/format/showfull?sysnb=0117836}.

\bibitem{Rajan:2017lxk}
R.~Gupta, Y.-C. Jang, H.-W. Lin, B.~Yoon and T.~Bhattacharya, \emph{{Axial
  Vector Form Factors of the Nucleon from Lattice QCD}},
  \href{https://doi.org/10.1103/PhysRevD.96.114503}{\emph{Phys. Rev.}
  {\bfseries D96} (2017) 114503}
  [\href{https://arxiv.org/abs/1705.06834}{{\ttfamily 1705.06834}}].

\bibitem{Bar:2018akl}
O.~B{\"a}r, \emph{{Nucleon-pion-state contamination in lattice calculations of
  the axial form factors of the nucleon}},  in \emph{{36th International
  Symposium on Lattice Field Theory (Lattice 2018) East Lansing, MI, United
  States, July 22-28, 2018}}, 2018,
  \href{https://arxiv.org/abs/1808.08738}{{\ttfamily 1808.08738}}.

\bibitem{Bar:2018xyi}
O.~B{\"a}r, \emph{{$N\pi$-state contamination in lattice calculations of the
  nucleon axial form factors}},
  \href{https://doi.org/10.1103/PhysRevD.99.054506}{\emph{Phys. Rev. D}
  {\bfseries 99} (2019) 054506}
  [\href{https://arxiv.org/abs/1812.09191}{{\ttfamily 1812.09191}}].

\bibitem{Jang:2019vkm}
Y.-C. Jang, R.~Gupta, B.~Yoon and T.~Bhattacharya, \emph{{Axial Vector Form
  Factors from Lattice QCD that Satisfy the PCAC Relation}},
  \href{https://doi.org/10.1103/PhysRevLett.124.072002}{\emph{Phys. Rev. Lett.}
  {\bfseries 124} (2020) 072002}
  [\href{https://arxiv.org/abs/1905.06470}{{\ttfamily 1905.06470}}].

\bibitem{Lee:2015jqa}
G.~Lee, J.~R. Arrington and R.~J. Hill, \emph{{Extraction of the proton radius
  from electron-proton scattering data}},
  \href{https://doi.org/10.1103/PhysRevD.92.013013}{\emph{Phys. Rev. D}
  {\bfseries 92} (2015) 013013}
  [\href{https://arxiv.org/abs/1505.01489}{{\ttfamily 1505.01489}}].

\bibitem{Meyer:2021vfq}
A.~S. Meyer et~al., \emph{{Nucleon Axial Form Factor from Domain Wall on
  HISQ}}, \href{https://doi.org/10.22323/1.396.0081}{\emph{PoS} {\bfseries
  LATTICE2021} (2022) 081} [\href{https://arxiv.org/abs/2111.06333}{{\ttfamily
  2111.06333}}].

\bibitem{Tsuji:2023llh}
{\scshape PACS} collaboration, R.~Tsuji, Y.~Aoki, K.-I. Ishikawa, Y.~Kuramashi,
  S.~Sasaki, K.~Sato et~al., \emph{{Nucleon form factors in $N_f=2+1$ lattice
  QCD at the physical point : finite lattice spacing effect on the
  root-mean-square radii}},  \href{https://arxiv.org/abs/2311.10345}{{\ttfamily
  2311.10345}}.

\bibitem{Ohta:2023ygq}
{\scshape LHP, RBC,, UKQCD} collaboration, S.~Ohta, \emph{{Nucleon isovector
  form factors from domain-wall lattice QCD at the physical mass}},  in
  \emph{{40th International Symposium on Lattice Field Theory}}, 11, 2023,
  \href{https://arxiv.org/abs/2311.05894}{{\ttfamily 2311.05894}}.

\bibitem{Chang:2018uxx}
{\scshape CalLat} collaboration, C.~C. Chang et~al., \emph{{A per-cent-level
  determination of the nucleon axial coupling from quantum chromodynamics}},
  \href{https://doi.org/10.1038/s41586-018-0161-8}{\emph{Nature} (2018) }
  [\href{https://arxiv.org/abs/1805.12130}{{\ttfamily 1805.12130}}].

\bibitem{Gupta:2018qil}
R.~Gupta, Y.-C. Jang, B.~Yoon, H.-W. Lin, V.~Cirigliano and T.~Bhattacharya,
  \emph{{Isovector Charges of the Nucleon from 2+1+1-flavor Lattice QCD}},
  \href{https://doi.org/10.1103/PhysRevD.98.034503}{\emph{Phys. Rev.}
  {\bfseries D98} (2018) 034503}
  [\href{https://arxiv.org/abs/1806.09006}{{\ttfamily 1806.09006}}].

\bibitem{Harris:2019bih}
T.~Harris, G.~von Hippel, P.~Junnarkar, H.~B. Meyer, K.~Ottnad, J.~Wilhelm
  et~al., \emph{{Nucleon isovector charges and twist-2 matrix elements with
  $N_f=2+1$ dynamical Wilson quarks}},
  \href{https://doi.org/10.1103/PhysRevD.100.034513}{\emph{Phys. Rev. D}
  {\bfseries 100} (2019) 034513}
  [\href{https://arxiv.org/abs/1905.01291}{{\ttfamily 1905.01291}}].

\bibitem{Liang:2018pis}
{[$\chi$QCD 18] J. Liang}, Y.-B. Yang, T.~Draper, M.~Gong and K.-F. Liu,
  \emph{{Quark spins and Anomalous Ward Identity}},
  \href{https://doi.org/10.1103/PhysRevD.98.074505}{\emph{Phys. Rev.}
  {\bfseries D98} (2018) 074505}
  [\href{https://arxiv.org/abs/1806.08366}{{\ttfamily 1806.08366}}].

\bibitem{Irani:2023lol}
F.~Irani, M.~Goharipour, H.~Hashamipour and K.~Azizi, \emph{{New insight on the
  nucleon structure from recent MINERvA measurement of the antineutrino-proton
  scattering cross-section}},  6, 2023.

\bibitem{Bali:2016lva}
G.~S. Bali, B.~Lang, B.~U. Musch and A.~Sch\"afer, \emph{{Novel quark smearing
  for hadrons with high momenta in lattice QCD}},
  \href{https://doi.org/10.1103/PhysRevD.93.094515}{\emph{Phys. Rev. D}
  {\bfseries 93} (2016) 094515}
  [\href{https://arxiv.org/abs/1602.05525}{{\ttfamily 1602.05525}}].

\bibitem{Edwards:2004sx}
{\scshape SciDAC Collaboration, LHPC Collaboration, UKQCD Collaboration}
  collaboration, R.~G. Edwards and B.~Joo, \emph{{The Chroma software system
  for lattice QCD}},
  \href{https://doi.org/10.1016/j.nuclphysbps.2004.11.254}{\emph{Nucl.Phys.Proc.Suppl.}
  {\bfseries 140} (2005) 832}
  [\href{https://arxiv.org/abs/hep-lat/0409003}{{\ttfamily hep-lat/0409003}}].

\end{thebibliography}\endgroup
